\documentclass[titlepage,12pt]{utarticle}
\usepackage{amsmath,amsfonts,bbm,mathrsfs}
\usepackage[dvips]{graphics,graphicx,psfrag}
\usepackage{epsfig}
\usepackage{color}
\numberwithin{equation}{section}

\begin{document}

\preprint{
  CERN--PH--TH/2007-021\\
  {\tt hep-th/0701269}\\
}

\title{\vskip-1cm Matrix Factorizations and Homological Mirror Symmetry on the Torus}
\author{Johanna Knapp \and Harun Omer}
\oneaddress{Department of Physics, CERN\\
     Theory Division,
     CH-1211 Geneva 23,
     Switzerland\\
    {\tt firstname.lastname@cern.ch}\\{~}\\}

 \nobreak
\Abstract{We consider matrix factorizations and homological mirror symmetry on the torus $T^2$ using a Landau--Ginzburg description. We identify the basic matrix factorizations of the Landau--Ginzburg superpotential and compute the full spectrum, taking into account the explicit dependence on bulk and boundary moduli. We verify homological mirror symmetry by comparing three--point functions in the A--model and the B--model.
}

\date{January 2007}
\maketitle
\tableofcontents
\pagebreak
\section{Introduction}
\label{intro-sec}
Matrix factorizations of Landau--Ginzburg superpotentials provide a practical
tool to describe D--branes in B--type topological string theory.
D--branes are often imagined as objects wrapping cycles on a
Calabi--Yau manifold. This is of course only an appropriate
concept in special limits, away from which this kind of intuitive
understanding breaks down. A more appropriate framework for the
description of topological $D$--branes is in terms of Fukaya categories
(on the $A$ side) or derived categories of coherent sheaves (on the
$B$ model side)~\cite{Sharpe:1999qz,Douglas:2000gi,Lazaroiu:2001jm,Aspinwall:2001pu,
Lazaroiu:2003md,orlov1,Orlov:2003yp,Kapustin:2002bi,mirrorsymmetry}.
Brane stability, brane composition through
tachyon condensation, brane locations as well as brane charges
are all included in this framework.
The formulation is rather abstract and still relatively new, so the number
of concrete calculations done which have been done explicitly is still very small. 
The $B$--side whose Landau--Ginzburg description realizes $D$--branes in terms of matrix factorizations
has to date been discussed for minimal models as well as for tensor products
of them~\cite{Brunner:2003dc,Kapustin:2003rc,Ashok:2004zb,Brunner:2005fv}.
The only explicit calculation for a moduli dependent problem
was performed for the torus with Landau--Ginzburg desciption $T^2/\mathbbm{Z}_3$. The elliptic
curve's mirror description on the $A$--side has been studied 
from the physical as well as the mathematical perspective
in~\cite{Brunner:2004mt,Herbst:2006nn,Govindarajan:2005im,Govindarajan:2006uy}
and~\cite{Polishchuk:1998db,PolishchukFukaya,Polishchuk:2000kx} respectively.
\\
Three different Landau--Ginzburg orbifolds of $T^2$ are known.
The $\mathbbm{Z}_3$ orbifold is described by a cubic LG superpotential
and the $\mathbbm{Z}_4$ orbifold by a quartic curve. The third superpotential listed
below corresponds to the $\mathbb{Z}_6$ orbifold. Explicitly, the
associated Landau--Ginzburg superpotentials take the form (see e.g. \cite{Lerche:1989cs}):
\begin{align}
\label{cubic}
W_{\mathbb{Z}_3}&=x_1^3+x_2^3\:+x_3^3-a\:x_1x_2x_3\\
\label{quartic}
W_{\mathbb{Z}_4}&=x_1^4+x_2^4\:-x_3^2-a\:x_1^2x_2^2\\
W_{\mathbb{Z}_6}&=x_1^6+x_2^3\:- x_3^2-a\:x_1x_2^4.
\end{align}
The parameter $a$ depends on the complex structure modulus $\tau$ of the torus.
The Gepner point corresponds to $a=0$. These models correspond to different points in the K\"ahler moduli space.  Adding a quadratic term
$x_3^2$ to the above superpotentials has no effect on the bulk theory but for a theory with boundary this
is different. In a theory with boundary, adding a quadratic term to the superpotential gives a different GSO projection.\\
In this work we focus on the quartic curve (\ref{quartic}) which describes a Laundau--Ginzburg representation of the orbifold $T^2/\mathbbm{Z}_4$. We will often refer to this model as the quartic torus.
In the framework of matrix factorizations and derived categories the quartic curve
has so far only been treated at the Gepner point~\cite{Dell'Aquila:2005jg,Schmidt-Colinet:2007vi}.
Here we will discuss the complete, moduli--dependent quartic torus without
restricting to any special point like the Gepner point. Therefore, all results
depend on the bulk modulus $\tau$ as well as boundary moduli $u_i$.
Many of the methods we apply here were developed for the case of the cubic curve.\\\\
By working with the $T^2$ we have deliberately chosen a case whose physics and mathematics is already
well--understood. We hope that this paper is useful to allow tackling physically more interesting cases, notably branes on the $K3$, 
semi-realistic intersecting brane models on higher dimensional tori, or branes on higher dimensional Calabi--Yaus.
The aforementioned cases are of great theoretical interest but knowledge about them
is only spotty since they are tractable only to a low level of sophistication by conventional
methods; it is hoped that a matrix factorization perspective can reveal new insights.
Apart from the explicit moduli--dependence, matrix factorizations allow
finding in principle all possible stable brane configurations. 
In addition, compared to the cubic curve, the quartic curve has a few novel features.
The most obvious is that the variables of the quartic superpotential do not have equal weights. This has an influence
on the uniformization procedure which relates the moduli dependent parameters
of the B--model to those natural in the A--model.
Another difference between the quartic and the cubic torus is that the superpotential
contains only terms with even exponents. This implies that there will be a
self-dual matrix factorization, i.e. a brane which is its own antibrane.
The existence of self-dual matrix factorizations then entails that we must include
antibranes into the description if we want to compute correlators on the quartic torus.
Furthermore, selection rules tell us that the three--point functions for the 
quartic torus always have insertions of two different kinds of factorizations whereas
for the cubic curve all calculation could be done with just one type.
Throughout this paper we will distinguish between the two types of factorizations,
calling them 'long' and 'short' branes, in line with previous conventions. In the A--model picture the
long branes wrap along the diagonal of the fundamental domain of the torus, whereas the short branes wrap along the sides of the
covering space. On the cubic, there were two kinds of three--point
correlators which involved only states stretching between long branes or short only branes, respectively.
In the case of the quartic torus, the three--point correlators always involve long as well as short branes.\\\\
The paper is organized as follows: We start by identifying a set of matrix
factorizations -- the long and short branes in section \ref{mf-sec} -- from which all other branes
can be generated by tachyon condensation \cite{Govindarajan:2005im}.
In section \ref{cohsec} we compute the complete spectrum of the model. Section~\ref{bcorr-sec} is devoted to
the calculation of three--point functions on the B--model side. In that section we
also solve the uniformization problem, alluded above. Section~\ref{d2sec} is concerned with
an exceptional matrix factorization, which corresponds to a pure D2 brane wrapping
the torus. Its boundary modulus is therefore fixed. We will be somewhat elaborate
on this subject since this factorization is important, for example, for computing
monodromies in moduli space~\cite{Jockers:2006sm}.  In section~\ref{acorr-sec} we compute the
instanton sums on the A--model side and verify that they match the B--model
calculation. Furthermore we present some examples of higher point amplitudes. In section~\ref{twovarsec}
we discuss another version of the quartic curve, where $x_3=0$. In this
two--variable version of the model the boundary modulus is set to a fixed value
and thus we work with a theory that depends only on the complex structure modulus $\tau$.
The main goal of this section is to show that for a fixed boundary modulus we can avoid the uniformization procedure by extending the formalism for
the construction of permutation branes \cite{Brunner:2005fv,Enger:2005jk}.
Using this trick, we can take over many of the results of the CFT constructions. The aim of this section is to point out the generalization of the CFT constructions for this case.
Finally, in the appendix, we give the boundary--changing spectrum of the quartic torus as well as
some definitions and identities for theta functions which were used.
\section{Matrix Factorizations}
\label{mf-sec}
We consider the following three--variable Landau--Ginzburg superpotential for the quartic torus:
\begin{equation}
\label{supo}
W=x_1^4+x_2^4-x_3^2-ax_1^2x_2^2
\end{equation}
In order to incorporate moduli in a natural manner we introduce
parameters $\alpha_1^i\equiv\alpha_1(u_i,\tau)$, $\alpha_2^i\equiv\alpha_2(u_i,\tau)$,
which depend on the boundary modulus $u$ of the brane (we label the brane by an index $i$) and
the complex structure modulus $\tau$ of the torus. The matrix factorization condition constrains
the $\alpha_i$ to lie on the Jacobian of the torus. The parameters therefore have to satisfy the
following relation:
\begin{equation}
\label{alpharel}
(\alpha_1^i)^4+(\alpha_2^i)^4-(\alpha_3^i)^2-a(\alpha_1^i)^2(\alpha_2^i)^2=0
\end{equation}
We now give the matrix factorizations which correspond to the long and short branes in the $A$--model. It is a priori not obvious to see from the structure of the factorization whether it is a long brane or a short brane. One way to find out is by computing the RR--charges \cite{Walcher:2004tx,Govindarajan:2005im}. We choose a different  approach, identifying the branes by their spectra. From the A--model picture we know that long branes intersect twice in the fundamental domain of the torus, which implies that we have two states stretching between two long branes. Short branes intersect only once and we expect one open string state between two short branes. We will show explicitly in the following section and in the appendix that the matrix factorizations given below satisfy these properties.\\
For the brane anti--brane pair of the short branes we find: $Q^{\mathcal{S}}_i=\left(\begin{array}{cc}0&E_i^{\mathcal{S}}\\J_i^{\mathcal{S}}&0\end{array}\right)$, where $i=\{1,\ldots,4\}$:
\begin{equation}
\label{short1}
E_i^{\mathcal{S}}=\left(\begin{array}{cc}
\alpha_1^ix_1+\alpha_2^ix_2&\alpha_3^ix_3+\frac{(\alpha_3^i)^2}{\alpha_1^i\alpha_2^i}x_1x_2\\
\frac{1}{\alpha_1^i\alpha_2^i}x_1x_2-\frac{1}{\alpha_3^i}x_3&-\frac{1}{\alpha_1^i}x_1^3+\frac{\alpha_1^i}{(\alpha_2^i)^2}x_1x_2^2-\frac{1}{\alpha_2^i}x_2^3+\frac{\alpha_2^i}{(\alpha_1^i)^2}x_1^2x_2
\end{array}\right)
\end{equation} 
\begin{equation}
\label{short2}
J_i^{\mathcal{S}}=\left(\begin{array}{cc}
\frac{1}{\alpha_1^i}x_1^3-\frac{\alpha_1^i}{(\alpha_2^i)^2}x_1x_2^2+\frac{1}{\alpha_2^i}x_2^3-\frac{\alpha_2^i}{(\alpha_1^i)^2}x_1^2x_2&\alpha_3^ix_3+\frac{(\alpha_3^i)^2}{\alpha_1^i\alpha_2^i}x_1x_2\\
\frac{1}{\alpha_1^i\alpha_2^i}x_1x_2-\frac{1}{\alpha_3^i}x_3&-\alpha_1^ix_1-\alpha_2^ix_2
\end{array}\right)
\end{equation}
This has the structure of a linear permutation brane. This is in accordance with the case of the cubic curve \cite{Brunner:2004mt}, where the short branes were also identified with linear permutation branes.\\
For the long branes we take the following expression: 
$Q^{\mathcal{L}}_i=\left(\begin{array}{cc}0&E_i^{\mathcal{L}}\\J_i^{\mathcal{L}}&0\end{array}\right)$, where:
\begin{equation}
\label{long1}
E_i^{\mathcal{L}}=\left(\begin{array}{cc}
\frac{\alpha_1^i}{\alpha_2^i}x_1^2-\frac{\alpha_2^i}{\alpha_1^i}x_2^2&\frac{1}{\alpha_3^i}x_3-\frac{1}{\alpha_1^i\alpha_2^i}x_1x_2\\
\alpha_3^ix_3+\frac{(\alpha_3^i)^2}{\alpha_1^i\alpha_2^i}x_1x_2&\frac{\alpha_2^i}{\alpha_1^i}x_1^2-\frac{\alpha_1^i}{\alpha_2^i}x_2^2
\end{array}\right)
\end{equation}
\begin{equation}
\label{long2}
J_i^{\mathcal{L}}=\left(\begin{array}{cc}
\frac{\alpha_2^i}{\alpha_1^i}x_1^2-\frac{\alpha_1^i}{\alpha_2^i}x_2^2&-\frac{1}{\alpha_3^i}x_3+\frac{1}{\alpha_1^i\alpha_2^i}x_1x_2\\
-\alpha_3^ix_3-\frac{(\alpha_3^i)^2}{\alpha_1^i\alpha_2^i}x_1x_2&\frac{\alpha_1^i}{\alpha_2^i}x_1^2-\frac{\alpha_2^i}{\alpha_1^i}x_2^2
\end{array}\right)
\end{equation}
These matrix factorizations correspond to Recknagel--Schomerus branes. \\
Note that, apart from the permutation branes, there is a standard construction for a matrix factorization. One can factorize the superpotential as follows: $W=\sum_i q_i x_i\frac{\partial W}{\partial x_i}$. In our case this would yield a $4\times 4$ matrix factorization. Comparing with the cubic curve, one might expect that a factorization of this kind would give the long branes. We will argue in section~\ref{d2sec} why~(\ref{long1}-\ref{long2}) is the simplest choice for the long branes.\\
Since we have a $\mathbb{Z}_4$--orbifold action the index $i$ can take the values $i\in\{1,2,3,4\}$. The R--matrices are given by:
\begin{eqnarray}
R_1&=&\mathrm{diag}\left(\frac{1}{4},-\frac{1}{4},-\frac{1}{4},\frac{1}{4}\right)\\
R_2&=&\mathrm{diag}(0,0,0,0)
\end{eqnarray}
The orbifold matrices associated to the matrix factorizations above are \cite{Walcher:2004tx}:
\begin{equation}
\gamma_{1,2}^i=\sigma e^{i\pi R_{1,2}}e^{-i\pi\varphi_i},
\end{equation}
where $\sigma=\left(\begin{array}{cc}\mathbbm{1}_4&0\\0&-\mathbbm{1}_4\end{array}\right)$ and the phase $\varphi$ is determined by the condition $\gamma^4=\mathbbm{1}$.

\section{Cohomology}
\label{cohsec}
\subsection{R--charges}
\begin{figure}
\begin{center}
\includegraphics{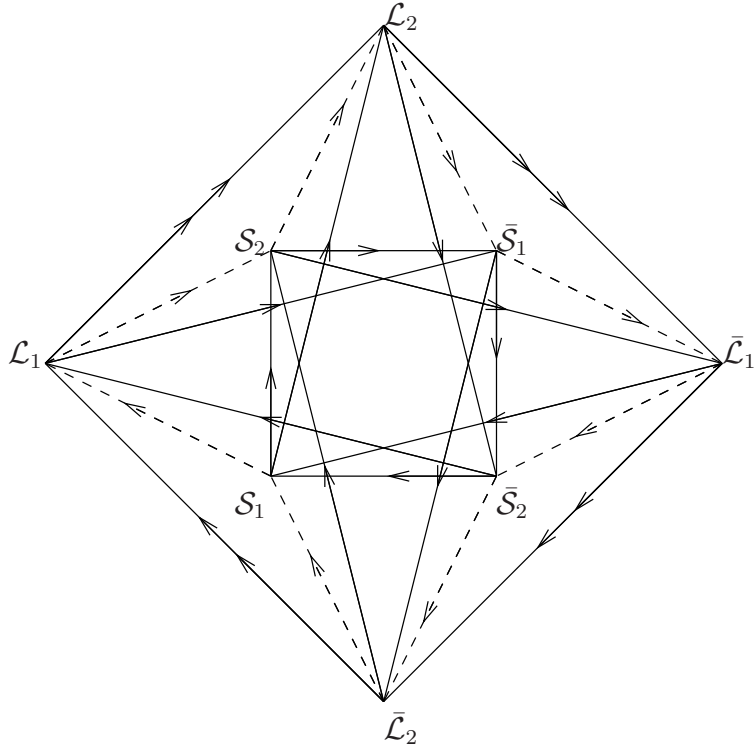}
\end{center}
\caption{The quiver diagram for the quartic torus.}\label{bquartic1}
\end{figure}

\begin{figure}
\begin{center}
\includegraphics{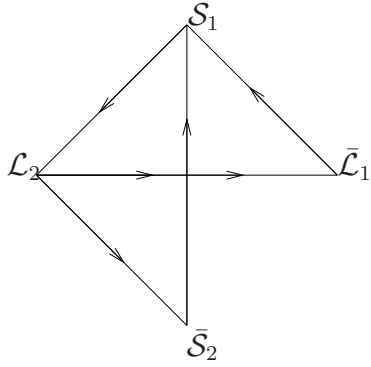}
\end{center}
\caption{Three--point functions on the B--side.}\label{bquartic2}
\end{figure}

\begin{figure}
\begin{center}
\includegraphics{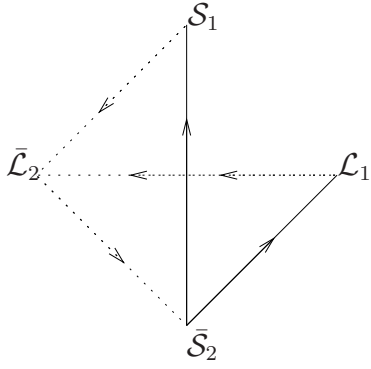}
\end{center}
\caption{Three--point functions on the B--side, including bosons.}\label{bquartic3}
\end{figure}

Here some basic results are derived about $R$--charges, which are used
throughout this work. For practical computations with branes described
by matrix factorizations, it is essential to know about the open string
states stretching between the
branes. Finding these states is a cohomology problem which can be
solved algebraically. Though not challenging from the theoretical
point of view, the computation can be cumbersome, especially 
for moduli dependent problems. From a phenomenological perspective,
it would therefore be helpful to extract some more information
from the factorization before actually committing to compute the cohomology.
Viewed from the practical side, it is easier to find the matrices in the
cohomology if we know those which can exist beforehand
and know the charges of their matrix elements.\\
The $R$-charge $q_{\Phi}$ of a morphism $\Phi$ mapping
between $Q$ and $Q'$ is obtained from the equation~\cite{Walcher:2004tx},
\begin{equation}
E\Phi+R'\Phi-\Phi R=q_{\Phi}\Phi.
\end{equation}
Without explicitly knowing the morphism itself, that condition allows to
determine the $R$-charges of the components from the (diagonal) $R$-matrices
$R=\mbox{diag}(r_1,r_2,...,r_m)$ and $R'=\mbox{diag}(r'_1,r'_2,...,r'_n)$
associated with the two branes. The charges of each entry therefore satisfy,
\begin{equation}
ch(\Phi_{ij})+r'_i-r_j=q_\Phi \qquad \mbox{or  }\Phi_{ij}=0. \label{eq:charge}
\end{equation}
A further obvious condition is that since the entries of the matrix
$\Phi$ are all proportional to polynomials in the chiral
ring $\mathcal{J}=\frac{\mathbb{C}[x_1,...,x_N]}{\partial W}$, they
can also only assume the discrete charges of these chiral ring elements,
\begin{equation}
ch(\Phi_{ij}) \in \{ch(x)|x\in \mathcal{J} \}.
\end{equation}
A third restriction is Serre duality: On a Calabi-Yau,
for every boson $\phi$ there exists a fermion $\psi$ and the charges of
both add up to the background charge $\hat c$.
\begin{equation}
q_\phi+q_\psi=\hat c. \qquad(0<q_\phi,q_\psi< \hat c) \label{eq:serre}
\end{equation}
Finally, there is the orbifold condition~\cite{Walcher:2004tx},
\begin{equation}
q_\Phi=\varphi'-\varphi+|\Phi| \mbox{ mod }2.\label{eq:orbi}
\end{equation}
In that equation, $|\Phi|$ denotes the parity of the morphism and
$\varphi$ is the phase of the matrix factorization. This phase is defined by,
\begin{eqnarray}
\gamma=\mbox{diag}(\mathbbm{1}_{N\times N},-\mathbbm{1}_{N\times N}) e^{i \pi R} e^{-i \pi \varphi},
\qquad \gamma^{H}=\mathbbm{1},
\end{eqnarray}
where $H$ is the total degree of the polynomial Landau-Ginzburg superpotential
$W_{LG}$.\\
Using these rules we can make ans\"atze for the physical states. 
Computing the boundary changing open string spectrum between the branes Eq.~(\ref{short1}--\ref{long2}), one obtains the quiver diagram depicted in Fig. \ref{bquartic1}.
\psfrag{l2}{$\mathcal{L}_2$}
\psfrag{bs1}{$\bar{\mathcal{S}}_1$}
\psfrag{bl1}{$\bar{\mathcal{L}}_1$}
\psfrag{bs2}{$\bar{\mathcal{S}}_2$}
\psfrag{bl2}{$\bar{\mathcal{L}}_2$}
\psfrag{s2}{$\mathcal{S}_2$}
\psfrag{l1}{$\mathcal{L}_1$}
\psfrag{s1}{$\mathcal{S}_1$}
Here, only the fermionic states have been drawn. By Serre duality, the bosonic states run in the opposite direction. The torus has background charge~$\hat c=1$ so~\ref{eq:serre} becomes $q_{\psi}+q_{\phi}=1$. The states mapping between branes of the same type have charge $1/2$. This tells us that at a generic point in moduli space there will be no non--vanishing three--point functions if one considers only short branes or only long branes. For the open string states between long and short branes the solid lines represent fermions of charge $1/4$ and the dashed lines are fermions of charge $3/4$.\\
For every brane, there is also a fermionic boundary preserving operator of charge 1, which corresponds to a marginal boundary deformation. It is given by $\Omega=\partial_u Q$, where u is the boundary modulus. This state will yield a one--point function $\langle\Omega\rangle$ for each of the long and short branes.\\
Furthermore, there exist additional states between a brane and its antibrane if the branes lie on top of each other. In particular, there will be a boson of charge 1 and a fermion of charge 0. The existence of these states comes from the fact that a boson, resp. fermion, beginning and ending on the same brane implies the existence of a fermion, resp. boson, stretching between the brane and its antibrane. In this sense, these states are related to $\mathbbm{1}$ and $\Omega$. We will not pursue this degenerate case any further in this work.\\\\
In the following, we will be interested in computing the non--vanishing three--point functions for the quartic torus. Clearly, only the fermions with charge $1/4$ and $1/2$ can contribute to the three--point functions. The possible three--point correlators correspond to oriented triangles in figure \ref{bquartic1}.  Note that we did not draw the bosons in this graph. In order to identify all the three-point functions one has to keep in mind that there is also a bosonic arrow going in the opposite direction. The quiver in fig. 
\ref{bquartic1} has an obvious $\mathbb{Z}_4$ symmetry. So, each type of correlator appears four times. \\
Let us first show some examples. We cut out a patch of Fig.~\ref{bquartic1} which contains all the information about correlators with only fermions. This is shown in Fig.~\ref{bquartic2}.
The possible correlators are thus:
\begin{equation}
\langle\psi_{\bar{\mathcal{S}}_2\mathcal{S}_1}\psi_{\mathcal{S}_1\mathcal{L}_2}\psi_{\mathcal{L}_2\bar{\mathcal{S}}_2}\rangle, \quad \langle\psi_{\bar{\mathcal{L}}_1\mathcal{S}_1}\psi_{\mathcal{S}_1\mathcal{L}_2}\psi_{\mathcal{L}_2\bar{\mathcal{L}}_1}\rangle,\quad \langle\psi_{\bar{\mathcal{L}}_1\mathcal{S}_1}\psi_{\mathcal{S}_1\mathcal{L}_2}\bar{\psi}_{\mathcal{L}_2\bar{\mathcal{L}}_1}\rangle.
\end{equation}
To find correlators with bosonic insertions, we have to swap the directions of some arrows in the quiver. One has, for instance, a configuration shown in Fig.~\ref{bquartic3} where bosons are represented by dotted lies.
Since the labeling of the branes is just convention and since we cannot tell the difference whether a state goes from brane to brane or from antibrane to antibrane in the B--model we will label the states in the correlators just with $\mathcal{L}$ and $\mathcal{S}$ and not with $\mathcal{L}_1,\bar{\mathcal{L}}_2,\bar{\mathcal{S}}_2$, etc.\\
There are two different types of correlators, those with two long branes and one short brane and those with two short branes and one long branes. In this paper we will mostly be concerned with the first type.\\
There are eight different correlators of type long--long--short:
\begin{equation}
\label{fercorr}
\langle\psi_{\mathcal{L}\mathcal{L}}\psi_{\mathcal{L}\mathcal{S}}\psi_{\mathcal{S}\mathcal{L}} \rangle\qquad \langle\bar{\psi}_{\mathcal{L}\mathcal{L}}\psi_{\mathcal{L}\mathcal{S}}\psi_{\mathcal{S}\mathcal{L}} \rangle
\end{equation}
\begin{equation}
\langle\psi_{\mathcal{L}\mathcal{L}}\phi_{\mathcal{L}\mathcal{S}}\phi_{\mathcal{S}\mathcal{L}} \rangle\qquad \langle\bar{\psi}_{\mathcal{L}\mathcal{L}}\phi_{\mathcal{L}\mathcal{S}}\phi_{\mathcal{S}\mathcal{L}} \rangle
\end{equation}
\begin{equation}
\langle\phi_{\mathcal{L}\mathcal{L}}\psi_{\mathcal{L}\mathcal{S}}\phi_{\mathcal{S}\mathcal{L}} \rangle\qquad \langle\bar{\phi}_{\mathcal{L}\mathcal{L}}\psi_{\mathcal{L}\mathcal{S}}\phi_{\mathcal{S}\mathcal{L}} \rangle
\end{equation}
\begin{equation}
\label{boscorr}
\langle\phi_{\mathcal{L}\mathcal{L}}\phi_{\mathcal{L}\mathcal{S}}\psi_{\mathcal{S}\mathcal{L}} \rangle\qquad \langle\bar{\phi}_{\mathcal{L}\mathcal{L}}\phi_{\mathcal{L}\mathcal{S}}\psi_{\mathcal{S}\mathcal{L}} \rangle
\end{equation}
Furthermore, there are four correlators of type short--short--long:
\begin{equation}
\langle\psi_{\mathcal{S}\mathcal{S}}\psi_{\mathcal{L}\mathcal{S}}\psi_{\mathcal{S}\mathcal{L}}\rangle \qquad \langle\psi_{\mathcal{S}\mathcal{S}}\phi_{\mathcal{L}\mathcal{S}}\phi_{\mathcal{S}\mathcal{L}}\rangle
\end{equation}
\begin{equation}
\langle\phi_{\mathcal{S}\mathcal{S}}\psi_{\mathcal{L}\mathcal{S}}\phi_{\mathcal{S}\mathcal{L}}\rangle \qquad \langle\phi_{\mathcal{S}\mathcal{S}}\phi_{\mathcal{L}\mathcal{S}}\psi_{\mathcal{S}\mathcal{L}}\rangle
\end{equation}
Here we used a bar to distinguish between the two states stretching between two long branes.\\
We collect the explicit results for all the open string states in the appendix.

\section{Three--point Functions in the B--model}
\label{bcorr-sec}
This section is concerned with the calculation of three--point correlators in the B--model. These can be determined by the following residue integral \cite{Kapustin:2003ga,Herbst:2004ax}:
\begin{equation}
\label{kapustin}
\langle\Phi_a^{AB}\Phi_b^{BC}\Phi_c^{CA}\rangle=\int\frac{\mathrm{str}\left(\frac{1}{3!}(\mathrm{d}Q^A)^{\wedge 3}\Phi^{AB}_a\Phi^{BC}_b\Phi^{CA}_c\right)}{\partial_1W\partial_2W\partial_3W},
\end{equation}
where the $\Phi_i^{AB}$ are (bosonic or fermionic) cohomology elements stretching between the branes $A$ and $B$.\\
The key difficulty in this computation is to find the correct normalization for the cohomology elements such that the correlators can be identified with the instanton sums in the A--model. This is related to finding flat coordinates on the moduli space. For the boundary preserving operator $\Omega=\partial_u Q$ the correct normalization can be deduced from the normalization of the superpotential which has to be normalized by a flattening normalization factor \cite{Lerche:1991wm,Brunner:2004mt}. The normalization of the boundary changing operators is more difficult to calculate. In \cite{Brunner:2004mt} it was argued that the correctly normalized three--point functions have to satisfy the heat equation, which implies that they have to be theta functions.\\
In the following subsections we will calculate the normalization for the boundary preserving operator $\Omega$ and compute the one--point functions. In order to be able to compute three--point functions, we first uniformize the moduli--dependent parameters $\alpha_i$, i.e. we express them in terms of the boundary modulus $u$ and the complex structure modulus $\tau$. Then we proceed to calculating the three--point functions in the B--model, making extensive use of theta function identities. 
\subsection{The flattening normalization factor}
The normalization of the boundary preserving operator $\Omega=\partial_u Q$ is related to the normalization of the superpotential $W$ via the matrix factorization condition. In \cite{Lerche:1991wm} it was shown that it is necessary to change the normalization of the superpotential by a modulus dependent prefactor, $W\rightarrow \frac{1}{q(\tau)}W$, in order to have vanishing connection terms in the differential equations satisfied by the periods. The matrix factorization condition then implies that we need to redefine $Q\rightarrow q(\tau)^{-\frac{1}{2}}Q$. This additional factor is then inherited by $\Omega$. We will now calculate the flattening normalization factor for the quartic torus, following the steps in \cite{Lerche:1991wm}.\\
We start by defining the following integrals, which are related to periods of differential forms, given a superpotential with $n$ variables:
\begin{equation}
u_0=(-1)^{\lambda}\Gamma(\lambda)\int \frac{q(s)}{W^{\lambda}}\mathrm{d}x_1\wedge\ldots\wedge\mathrm{d}x_n
\end{equation} 
\begin{equation}
u_{\alpha}^{(\lambda)}=(-1)^{\lambda+1}\Gamma(\lambda+1)\int_{\gamma}\frac{\phi_{\alpha}(x_i,s)}{W^{\lambda+1}}\mathrm{d}x_1\wedge\ldots\wedge\mathrm{d}x_n
\end{equation}
Here, $q(s)$ is a function of the moduli and the flattening factor we are looking for, which is in generic, non--flat coordinates. $\phi_{\alpha}(x_i,s)$ is a bulk cohomology element, $\gamma$ is a homology cycle and $\Gamma(\lambda)$ is the gamma function. It can then be shown \cite{Lerche:1991wm} that $u_0$ satisfies the following differential equation:
\begin{equation}
\frac{\partial^2}{\partial s_i\partial s_j}u_0=C_{ij}^{\alpha}u_{\alpha}^{(\lambda+1)}+\Gamma_{ij}^k\left(\frac{\partial}{\partial s_k}\right)u_0,
\end{equation} 
where $C_{ij}^{\alpha}$ are the structure constants of the bulk chiral ring and $\Gamma_{ij}^k$ is the Gauss--Manin connection. Transforming the generic coordinates $s_i$ to flat coordinates $t_i$ implies a vanishing connection $\Gamma\equiv 0$. In particular, this condition leads to a differential equation which determines the flattening factor $q(t)$.\\
We will now specialize to the quartic torus. Thus, $W$ is given by (\ref{supo}) and $n=3$. Furthermore, we set for the modulus of the torus $t\equiv\tau$, as usual. Then we have $u_0=(-1)^{\lambda}\Gamma(\lambda)\int \frac{q(\tau)}{W^{\lambda}}\mathrm{d}x_1\wedge\mathrm{d}x_2\wedge\mathrm{d}x_3$. In the following we will drop the integral measure. Computing the second derivative of $u_0$ with respect to the modulus, one gets:
\begin{equation}
\frac{\partial^2 u_0}{\partial\tau^2}=\frac{q''}{q}u_0+(-1)^{\lambda+1}\Gamma(\lambda+1)\int\frac{1}{W^{\lambda+1}}\left(\frac{2q'}{q}+\frac{a''}{a'}\right)(-q\,a'\,x_1^2x_2^2)+(-1)^{\lambda+2}\Gamma(\lambda+2)\int\frac{q}{W^{\lambda+2}}(a')^2x_1^4x_2^4
\end{equation}
Next, we partially integrate the third term, applying the following identities:
\begin{equation}
(4-a^2)x_1^4x_2^4=x_1x_2^3\left(x_2\partial_{x_1}W+\frac{1}{2}a\,x_1\partial_{x_2}W\right)
\end{equation}
\begin{equation}
x_2\partial_{x_2}W=x_2\left(4x_2^3-2a\,x_1^2x_2\right)
\end{equation}
The vanishing of the connection corresponds to the vanishing of the terms proportional to $\frac{1}{W^{\lambda+1}}$. This leads to a differential equation for $q(\tau)$ in terms of $a(\tau)$:
\begin{equation}
(-1)^{\lambda+1}\Gamma(\lambda+1)\int\frac{1}{W^{\lambda+1}}\left(\frac{2q'}{q}+\frac{a''}{a'}+\frac{2aa'}{4-a^2}\right)(-q\,a'\,x^2y^2)\stackrel{!}{=}0
\end{equation}
The integration can be performed easily and we solve for $q(\tau)$:
\begin{equation}
q(\tau)=\left(\frac{4-a(\tau)^2}{a'(\tau)}\right)^{\frac{1}{2}}
\end{equation}
One also finds another useful identity:
\begin{equation}
\eta^6(\tau)=\frac{(-1)^{\frac{7}{4}}}{2\pi^{\frac{3}{2}}}\frac{1}{q^2(\tau)}=\frac{(-1)^{\frac{7}{4}}}{2\pi^{\frac{3}{2}}}\frac{a'(\tau)}{4-a^2(\tau)}
\end{equation}
\subsection{The correlators $\langle\Omega_{\mathcal{S}}\rangle$ and $\langle\Omega_{\mathcal{L}}\rangle$}
As argued above, the correctly normalized correlators look as follows \cite{Brunner:2004mt},
\begin{equation}
\langle\Omega_{\mathcal{S}/\mathcal{L}}\rangle=q(\tau)\int\frac{\mathrm{str}\left(\frac{1}{3!}(\mathrm{d}Q^{\mathcal{S}/\mathcal{L}})^{\wedge 3}\partial_u Q^{\mathcal{S}/\mathcal{L}}\right)}{\partial_1W\partial_2W\partial_3W}=\int\frac{f(\tau,u)H(x)}{\partial_1W\partial_2W\partial_3W}=f(u,\tau),
\end{equation}
where $H(x)=\frac{1}{12}\mathrm{det}\partial_i\partial_jW$ is the Hessian. Going to the patch $\alpha_2(u,\tau)=1$ and using the relation coming from the vanishing of the $u$--derivative of (\ref{alpharel}) in the selected patch,
\begin{equation}
0=\partial_u\:\left(\alpha_1(u,\tau)^4+1-\alpha_3(u,\tau)^2-a\:\alpha_1(u,\tau)^2\right),
\end{equation}
one finds that the two correlators take the same value:
\begin{equation}
\langle\Omega_{\mathcal{S}}\rangle=\langle\Omega_{\mathcal{L}}\rangle=f(u,\tau)=q(\tau)\frac{1}{2}\frac{\partial_u\alpha_3(u,\tau)}{2\alpha_1(u,\tau)^3-a(\tau)\alpha_1(u,\tau)}
\end{equation}
One can show that $\langle\Omega_{\mathcal{S}/\mathcal{L}}\rangle=1$ is satisfied if $u$ is a flat coordinate on the Jacobian \cite{Brunner:2004mt}. On the torus, the holomorphic one--form looks as follows:
\begin{equation}
\label{hol1form}
\eta=q(\tau)\int_{\mathcal{C}}\frac{\omega}{W}, \qquad\textrm{where}\quad\omega=\sum_{i=1}^{3}(-1)^ix_i\mathrm{d}x_1\wedge\widehat{\mathrm{d}x_i}\wedge\mathrm{d}x_3,
\end{equation}
and $\mathcal{C}$ is a contour winding around the hypersurface $W=0$. In our local patch, where $W=W(\alpha_1,1,
\alpha_3)$, we have $\omega=-\mathrm{d}\alpha_1\wedge\mathrm{d}\alpha_3$. We can solve this contour integral using the residue theorem: $\int_{\mathcal{C}}\frac{\mathrm{d}W}{W}=1$. This is due to the fact that $W$ is zero along the torus, which implies that $\frac{1}{W}$ has a first order pole on the hypersurface. From the relation $\mathrm{d}W=\sum_{i=1}^{3}\frac{\partial W}{\partial \alpha_i}\mathrm{d}\alpha_i\vert_{\alpha_2=1}$ we obtain:
\begin{equation}
\mathrm{d}\alpha_1=\frac{\mathrm{d}W-\frac{\partial W}{\partial\alpha_3}\mathrm{d}\alpha_3}{\frac{\partial W}{\partial \alpha_1}}
\end{equation}
Inserting this into (\ref{hol1form}) and using the residue formula, we get:
\begin{equation}
\eta=q(\tau)\frac{\mathrm{d}\alpha_3}{\partial_{\alpha_1}W(\alpha_1,1,\alpha_3)}=\frac{q(\tau)}{2}\frac{\mathrm{d}\alpha_3}{2\alpha_1(u,\tau)^3-a\alpha_1(u,\tau)}
\end{equation}
The solution of $f(u,\tau)=1$ is thus given by:
\begin{equation}
u=\int_{\infty}^{\alpha_3}\eta,
\end{equation}
which shows that this is a flat coordinate on the Jacobian.
\subsection{Uniformization of the $\alpha_i$}
We now give the explicit expression for the functions $\alpha_i(u,\tau)$ in terms of the boundary modulus $u$ and the complex structure modulus $\tau$. For the torus this amounts to computing the mirror map, i.e. we express the coordinates of the B--model in terms of flat coordinates which are the natural variables in the A--model. In the case of the torus these are the complex structure parameter $\tau$, which will be identified with the K\"ahler parameter on the mirror, and the brane positions $u_i$ which will be identified with shift-- and Wilson line moduli on the $A$--side. \\
The $\alpha_i$ will be expressed by theta functions which give a basis of global sections of line bundles on the torus. The theta functions with characteristics are defined as follows\footnote{We have collected the relevant definitions and identities in the appendix.}:
\begin{equation}
\Theta\left[\begin{array}{c}c_1\\c_2\end{array}\right](u,\tau)=\sum_{m\in\mathbb{Z}}q^{(m+c_1)^2/2}e^{2\pi i(u+c_2)(m+c_1)},
\end{equation}
where $q=e^{2\pi i \tau}$. According to~\cite{Polishchuk:1998db}, the $n$ functions $\theta[\frac{a}{n},0](nu,n\tau)$ with $a\in \mathbb{Z}/n\mathbb{Z}$ are the global sections of degree $n$ line bundles $L^n$. For $n=2$, the following relation holds: 
\begin{equation}
\label{2var}
\Theta^4\left[\begin{array}{c}0\\0\end{array}\right](0,2\tau)+\Theta^4\left[\begin{array}{c}\frac{1}{2}\\0\end{array}\right](0,2\tau)-a\:\Theta^2\left[\begin{array}{c}0\\0\end{array}\right](0,2\tau)\Theta^2\left[\begin{array}{c}\frac{1}{2}\\0\end{array}\right](0,2\tau)=0,
\end{equation}
where $a$ can be found, for instance, in \cite{Harvey:1988ur,Lerche:1989cs,Giveon:1990ay}. The $a$--parameter defines a map $a:H^+/\Gamma(2)\rightarrow \mathbb{CP}^1$ from the fundamental region of the modular group $\Gamma(2)$ ($H^+$ denotes the upper half plane) to the Riemann sphere, given by $\mathbb{CP}^1$. In terms of the modular invariant $j(\tau)=\frac{1}{q}+744+\ldots$ it is given by the following expression \cite{Giveon:1990ay}: 
\begin{equation}
\label{jrel}
j(\tau)=\frac{16(a^2+12)^3}{(a^2-4)^2}
\end{equation}
Here, one has to be careful to choose the correct branch of the solution for $a$. \\
Clearly, the relation (\ref{2var}) is not what we are looking for, because if we identify $\alpha_1=\Theta[0,0](0,2\tau)$ and $\alpha_2=\Theta[1/2,0](0,2\tau)$, the relation is $\alpha_1^4+\alpha_2^4-a\alpha_1^2\alpha_2^2=0$ instead of (\ref{alpharel}). This is the relation for the two--variable version of the quartic superpotential. One sees that (\ref{2var}) is only satisfied at a single point, $u=0$, in the brane moduli space. We will consider the two--variable model at $u=0$ in section \ref{twovarsec}, where we will show that keeping only the explicit dependence on $\tau$ simplifies the calculations tremendously.\\\\
For the three--variable quartic torus we need to uniformize the $\alpha_i$ to satisfy (\ref{alpharel}). The correct basis of theta function is given by the Jacobi theta functions:
\begin{eqnarray}
\Theta_1(u,\tau)\equiv\Theta\left[\begin{array}{c}\frac{1}{2}\\\frac{1}{2}\end{array}\right](u,\tau)\quad \Theta_2(u,\tau)\equiv\Theta\left[\begin{array}{c}\frac{1}{2}\\0\end{array}\right](u,\tau)\\
\Theta_3(u,\tau)\equiv\Theta\left[\begin{array}{c}0\\0\end{array}\right](u,\tau)\quad\Theta_4(u,\tau)\equiv\Theta\left[\begin{array}{c}0\\\frac{1}{2}\end{array}\right](u,\tau)
\end{eqnarray}
It turns out that the correct solution looks as follows:
\begin{equation}
\label{uniformization}
\alpha_1(u,\tau)=\Theta_1(2u,2\tau)\qquad \alpha_2(u,\tau)=\Theta_4(2u,2\tau)
\end{equation}
\begin{equation}
\alpha_3(u,\tau)=\frac{\Theta_4^2(2\tau)}{\Theta_2(2\tau)\Theta_3(2\tau)}\Theta_2(2u,2\tau)\Theta_3(2u,2\tau),
\end{equation}
where we define $\Theta_i(\tau)\equiv\Theta_i(0,\tau)$.\\
Furthermore we write the parameter $a$ as:
\begin{equation}
a=\frac{\Theta_2^4(2\tau)+\Theta_3^4(2\tau)}{\Theta_2^2(2\tau)\Theta_3^2(2\tau)}
\end{equation}
One can check that this expression also satisfies (\ref{jrel}).\\
We can now show that with these definitions (\ref{alpharel}) is satisfied. The most elegant way is to prove this analytically, using the following quadratic identities for the theta functions (see for example \cite{FarkasKra}):
\begin{eqnarray}
\Theta_3^2(u,\tau)\Theta_4^2(\tau)&=&\Theta_4^2(u,\tau)\Theta_3^2(\tau)-\Theta_1^2(u,\tau)\Theta_2^2(\tau)\nonumber\\
\Theta_2^2(u,\tau)\Theta_4^2(\tau)&=&\Theta_4^2(u,\tau)\Theta_2^2(\tau)-\Theta_1^2(u,\tau)\Theta_3^2(\tau)
\end{eqnarray}
Note that for the quartic curve the uniformization is slightly more complicated than for the cubic curve. This is due to the fact that not all the variables have the same weight. As a consequence, $\alpha_3(u,\tau)$, which has twice the weight of $\alpha_1(u,\tau)$ and $\alpha_2(u,\tau)$, has to be expressed as a composite of theta functions. It is to be expected that this is always the case whenever a quadratic term is added to the superpotential.
\subsection{Three--point correlators}
We now calculate the correlators (\ref{fercorr})-(\ref{boscorr}) using (\ref{kapustin}). 
Plugging the states into the residue formula, the actual value of the correlator will be multiplied by a rational function of the $\alpha_i$ which can be absorbed into the normalization of the states. Since this task is quite complicated we will simplify the problem stepwise:
\begin{itemize}
\item Simplify the open string states using theta function identities.
\item Insert these simplified states into (\ref{kapustin}) and make use of more identities for theta functions to identify the correlators.
\end{itemize}
Performing these steps, we will be able to extract the values of the correlators without knowing the exact normalization. In general, one needs to know the precise normalization in order to make the results comparable to the A--model. Without further input it cannot be determined. Our approach is to pull out factors $\alpha_{1/2}(u_i+u_j+u_k,\tau)$ from every contribution to the Hessian, which are expected to be the correct values for the correlators. We then verify that the results are correct by comparing the the results coming from the mirror calculation.
We will now give a more detailed description of the steps mentioned above.

\subsubsection*{Simplification of the open string states}
The open string states given above are quite complicated matrices whose entries contain sums of quotients of the $\alpha_i$. We can use the uniformization in terms of theta functions to simplify these expressions. In particular, we can apply theta function identities such that the $\alpha_i$--expressions give only one quotient instead of sums and we can pull out common factors which we may throw away, since the correlators are only defined up to factors in the $\alpha_i$. For the correlators it is thus possible to reduce the number of terms by a factor of 8. For this we made use of the addition formulas for the theta functions (see for instance \cite{Mumford1}). We collected the most important ones in the appendix. For the simplification of the states we used (\ref{simp1}), (\ref{simp2}).

\subsubsection*{The Correlators}
By plugging the simplified open string states into (\ref{kapustin}) and making further manipulations with theta function identities, we can pull out a factor of all terms in the supertrace which contribute to the Hessian. In that case we need to apply the most general identities for our theta functions. In particular, we make extensive use of~(\ref{mytheta}).\\

Here we compute the correlators which involve two long branes and one short brane. Applying the theta function identities, one sees that one term on the rhs of (\ref{mytheta}) always vanishes\footnote{This comes from $\Theta_1(0,\tau)=0$.}, leaving us with a term of the form $\Theta_{(1/4)}(u_1+u_2+u_3,\tau)$ and some factors coming from the bad normalization of the states. Thus, up to the normalization factor, we find the following results for the correlators:
\begin{align}
\langle\psi_{\mathcal{L}\mathcal{L}}(u_1,u_2)\psi_{\mathcal{L}\mathcal{S}}(u_2,u_3)\psi_{\mathcal{S}\mathcal{L}}(u_3,u_1)\rangle&\sim\Theta_4(2(u_1+u_2-u_3),2\tau)\nonumber \\
\langle\bar{\psi}_{\mathcal{L}\mathcal{L}}(u_1,u_2)\psi_{\mathcal{L}\mathcal{S}}(u_2,u_3)\psi_{\mathcal{S}\mathcal{L}}(u_3,u_1)\rangle&\sim\Theta_1(2(u_1+u_2-u_3),2\tau)
\label{eq:corr1}
\end{align}
\begin{align}
\langle \psi_{\mathcal{L}\mathcal{L}}(u_1,u_2)\phi_{\mathcal{L}\mathcal{S}}(u_2,u_3)\phi_{\mathcal{S}\mathcal{L}}(u_3,u_1)\rangle&\sim\Theta_4(2(u_1+u_2+u_3),2\tau)\nonumber\\
\langle \bar{\psi}_{\mathcal{L}\mathcal{L}}(u_1,u_2)\phi_{\mathcal{L}\mathcal{S}}(u_2,u_3)\phi_{\mathcal{S}\mathcal{L}}(u_3,u_1)\rangle&\sim\Theta_1(2(u_1+u_2+u_3),2\tau)
\end{align}
\begin{align}
\langle\phi_{\mathcal{L}\mathcal{L}}(u_1,u_2)\psi_{\mathcal{L}\mathcal{S}}(u_2,u_3)\phi_{\mathcal{S}\mathcal{L}}(u_3,u_1)\rangle&\sim\Theta_1(2(u_1-u_2+u_3),2\tau)\nonumber\\
\langle\bar{\phi}_{\mathcal{L}\mathcal{L}}(u_1,u_2)\psi_{\mathcal{L}\mathcal{S}}(u_2,u_3)\phi_{\mathcal{S}\mathcal{L}}(u_3,u_1)\rangle&\sim\Theta_4(2(u_1-u_2+u_3),2\tau)
\end{align}
\begin{align}
\langle\phi_{\mathcal{L}\mathcal{L}}(u_1,u_2)\phi_{\mathcal{L}\mathcal{S}}(u_2,u_3)\psi_{\mathcal{S}\mathcal{L}}(u_3,u_1)\rangle&\sim\Theta_1(2(u_1-u_2-u_3),2\tau)\nonumber \\
\langle\bar{\phi}_{\mathcal{L}\mathcal{L}}(u_1,u_2)\phi_{\mathcal{L}\mathcal{S}}(u_2,u_3)\psi_{\mathcal{S}\mathcal{L}}(u_3,u_1)\rangle&\sim\Theta_4(2(u_1-u_2-u_3),2\tau)
\label{eq:corr4}
\end{align}
The correlators always come in pairs. One correlator vanishes for $u_i=0$, the other one does not. In the A--model these correlators correspond to the two triangles that can be enclosed by two long branes and one short brane in the fundamental domain of the torus. Furthermore note that the pairs of correlators differ only by the relative signs of the $u_i$. The reason is that a fermion stretching between brane $\mathcal{A}$ and brane $\mathcal{B}$ corresponds to a boson stretching between $\mathcal{A}$ and the antibrane $\bar{\mathcal{B}}$. As a consequence bosons have opposite orientation as compared to fermions in the A--model picture which amounts to a relative sign changes in the $u_i$.\\
The calculation of the correlators involving two short branes and one long brane is more involved and seems to require the knowledge of the exact normalization of the states. We thus refrain from computing these here and refer to the A--model results.

\section{The ``exceptional'' D2 brane}
\label{d2sec}
As promised, we will now give an explanation why the matrix factorization (\ref{long1}), (\ref{long2}) gives a more convenient description for the long branes than the ``canonical'' factorization $W=\sum_i q_i x_i\frac{\partial W}{\partial x_i}$. This construction yields a $4\times 4$ matrix factorization $Q_i=\left(\begin{array}{cc}0&E_i\\J_i&0\end{array}\right)$ with:\footnote{Note that this factorization does not come exactly from $W=\sum_i q_i x_i\frac{\partial W}{\partial x_i}$. There is an additional term proportional to $x_1x_2$ in the entries with $x_3$. Altering the structure in this way does not change the properties of this factorization but it has the effect that the $\alpha_i$--dependent prefactors are just quotients of the $\alpha_i$ and not polynomials. This simplifies the calculations tremendously.}
\begin{equation}
E_i=\left(\begin{array}{cccc}
\alpha_1^ix_1&\alpha_2^ix_2&\alpha_3^ix_3+\frac{(\alpha_3^i)^2}{\alpha_1^i\alpha_2^i}x_1x_2&0\\
\frac{1}{\alpha_2^i}x_2^3-\frac{\alpha_2^i}{(\alpha_1^i)^2}x_1^2x_2&-\frac{1}{\alpha_1^i}x_1^3+\frac{\alpha_1^i}{(\alpha_2^i)^2}x_1x_2^2&0&\alpha_3^ix_3+\frac{(\alpha_3^i)^2}{\alpha_1^i\alpha_2^i}x_1x_2\\
\frac{1}{\alpha_3^i}x_3-\frac{1}{\alpha_1^i\alpha_2^i}x_1x_2&0&\frac{1}{\alpha_1^i}x_1^3-\frac{\alpha_1^i}{(\alpha_2^i)^2}x_1x_2^2&\alpha_2^ix_2\\
0&\frac{1}{\alpha_3^i}x_3-\frac{1}{\alpha_1^i\alpha_2^i}x_1x_2 &\frac{1}{\alpha_2^i}x_2^3-\frac{\alpha_2^i}{(\alpha_1^i)^2}x_1^2x_2&-\alpha_1^i x_1
\end{array}\right)
\end{equation}
\begin{equation}
J_i=\left(\begin{array}{cccc}
\frac{1}{\alpha_1^i}x_1^3-\frac{\alpha_1^i}{(\alpha_2^i)^2}x_1x_2^2&\alpha_2^ix_2&-\alpha_3^ix_3-\frac{(\alpha_3^i)^2}{\alpha_1^i\alpha_2^i}x_1x_2&0\\
\frac{1}{\alpha_2^i}x_2^3-\frac{\alpha_2^i}{(\alpha_1^i)^2}x_1^2x_2&-\alpha_1^ix_1&0&-\alpha_3^ix_3-\frac{(\alpha_3^i)^2}{\alpha_1^i\alpha_2^i}x_1x_2\\
-\frac{1}{\alpha_3^i}x_3+\frac{1}{\alpha_1^i\alpha_2^i}x_1x_2&0&\alpha_1^ix_1&\alpha_2^ix_2\\
0&-\frac{1}{\alpha_3^i}x_3+\frac{1}{\alpha_1^i\alpha_2^i}x_1x_2&\frac{1}{\alpha_2^i}x_2^3-\frac{\alpha_2^i}{(\alpha_1^i)^2}x_1^2x_2&-\frac{1}{\alpha_1^i}x_1^3+\frac{\alpha_1^i}{(\alpha_2^i)^2}x_1x_2^2
\end{array}\right)
\end{equation}
A straightforward calculation shows that this matrix factorization, together with the factorization (\ref{short1}), (\ref{short2}) for the short branes, yields the same spectrum as depicted in Fig. 1. Thus, this factorization also represents the long branes. However, there is a catch: Let us compute the correlator $\langle\Omega\rangle$ of the marginal boundary preserving operator $\Omega=\partial_uQ$. In order to do this, we insert it into the residue formula for the three--point function \cite{Kapustin:2003ga}, which for this case looks as follows:
\begin{equation}
\langle\Omega\rangle=\int\frac{\mathrm{str}\left(\frac{1}{3!}(\mathrm{d}Q)^{\wedge 3}\partial_u Q\right)}{\partial_1W\partial_2W\partial_3W}
\end{equation} 
In order to give something non--vanishing, the supertrace should be proportional to the Hessian. Inserting into this formula, one finds that the supertrace is identically 0. \\
We interpret this as follows: Since $\Omega$ is the derivative of $Q$ with respect to the boundary modulus the vanishing of this correlator implies that the boundary modulus for this matrix factorization has a fixed value. Such matrix factorizations have already been discussed in \cite{Hori:2004ja,Govindarajan:2005im}. They are interpreted as a single rigid $D2$ brane wrapping the torus.\\
Although these special points in moduli space are an interesting issue, we do not want to restrict ourselves to a specific value of the boundary modulus but rather find the most general expression for the long branes. The discussion in \cite{Hori:2004ja,Govindarajan:2005im} implies that we have to add a pair of $D0\overline{D0}$ branes to the system. Then one of the branes can move freely on the torus, while its antibrane remains fixed and thus we have restored the boundary modulus as the relative distance between the $D0$--brane and the $\overline{D0}$--brane. The results of \cite{Hori:2004ja,Govindarajan:2005im} tell us that this can be done perturbing this matrix with the marginal boundary fermion, which will yield a reducible matrix factorization. \\
In order to find an expression for this operator (which is not $\partial_uQ$ but an equivalent description) we go to the Gepner point. There, Landau--Ginzburg description of the torus is a tensor product of two $A_3$ minimal models and one (trivial) $A_1$ piece: $A_3(x_1)\otimes A_3(x_2)\otimes A_1(x_3)$. Our matrix factorization has the following form at the Gepner point (see also: \cite{Dell'Aquila:2005jg}): $Q_{gep}=\left(\begin{array}{cc}0&E_{gep}\\J_{gep}&0\end{array}\right)$:
\begin{equation}
E_{gep}=\left(\begin{array}{cccc}
\label{long-gep}
x_1&x_2&x_3&0\\
x_2^3&-x_1^3&0&x_3\\
x_3&0&x_1^3&x_2\\
0&x_3&x_2^3&-x_1
\end{array}\right)\qquad
J_{gep}=\left(\begin{array}{cccc}
x_1^3&x_2&-x_3&0\\
x_2^3&-x_1&0&-x_3\\
-x_3&0&x_1&x_2\\
0&-x_3&x_2^3&-x_3^3
\end{array}\right)
\end{equation}
There is a unique fermionic state of weight $1$, which is the tensor product of the highest weight fermions of the two $A_3$ minimal models. We identify this state with the state $\Omega$: $\Omega_{gep}=\left(\begin{array}{cc}0&\Omega^{(0)}_{gep}\\\Omega^{(1)}_{gep}&0\end{array}\right)$,where
\begin{equation}
\Omega^{(0)}_{gep}=\left(\begin{array}{cccc}
0&0&0&1\\
0&0&-x_1^2x_2^2&0\\
0&x_1^2&0&0\\
-x_2^2&0&0&0
\end{array}\right)\qquad
\Omega^{(1)}_{gep}\left(\begin{array}{cccc}
0&0&0&x_1^2\\
0&0&-x_2^2&0\\
0&1&0&0\\
-x_1^2x_2^2&0&0&0
\end{array}\right);
\end{equation}
Perturbing (\ref{long-gep}) with this state and turning the moduli back on we find the following reducible matrix factorization: $Q_i^{red}=\left(\begin{array}{cc}0&E_i^{red}\\J_i^{red}&0\end{array}\right)$, where
\begin{equation}
E_i^{red}=\left(\begin{array}{cccc}
x_1&(\alpha_2^i)^2x_2&\frac{1}{\alpha_3^1}x_3+\left(-\frac{1}{\alpha_1^i\alpha_2^i}+\frac{(\alpha_2^i)^3}{\alpha_1^i(\alpha_3^i)^2}\right)x_1x_2&\frac{\alpha_1^i}{\alpha_2^i}\\
(\alpha_2^i)^2x_2^3+\left(-\frac{(\alpha_2^i)^4}{(\alpha_1^i)^2}+\frac{(\alpha_3^i)^2}{(\alpha_1^i)^2}\right)x_1^2x_2&-(\alpha_3^i)^2x_1^3&-\frac{\alpha_1^i}{\alpha_2^i}x_1^2x_2^2&\alpha_3^ix_3\\
\frac{1}{\alpha_3^i}x_3&\frac{\alpha_1^i}{\alpha_2^i}x_1^2&\frac{1}{(\alpha_3^i)^2}x_1^3&\frac{1}{(\alpha_2^i)^2}x_2\\
-\frac{\alpha_1^i}{\alpha_2^i}x_2^2&\mu_1&\mu_2&-x_1
\end{array}\right),
\end{equation}
where $\mu_1=\alpha_3^ix_3+\left(-\frac{(\alpha_2^i)^3}{\alpha_1^i}+
\frac{(\alpha_3^i)^2}{\alpha_1^i\alpha_2^i}\right)x_1x_2$ and $\mu_2=\frac{1}{(\alpha_2^i)^2}x_2^3+\left(\frac{1}{(\alpha_1^i)^2}-\frac{(\alpha_2^i)^4}{(\alpha_1^i)^2(\alpha_3^i)^2}\right)x_1^2x_2$.
\begin{equation}
J_i^{red}=\left(\begin{array}{cccc}
x_1^3&\frac{1}{(\alpha_2^i)^2}&-\alpha_3^1x_3&\frac{\alpha_1^i}{\alpha_2^i}x_1^2\\
\mu_2&-\frac{1}{(\alpha_3^1)^3}x_1&-\frac{\alpha_1^i}{\alpha_2^1}&-\frac{1}{\alpha_3^1}x_3-\left(-\frac{1}{\alpha_1^i\alpha_2^i}+\frac{(\alpha_2^i)^3}{\alpha_1^i(\alpha_3^i)^2}\right)x_1x_2\\
\mu_1&\frac{\alpha_1^i}{\alpha_2^i}&(\alpha_3^i)^2x_1&(\alpha_2^i)^2x_2\\
-\frac{\alpha_1^i}{\alpha_2^i}x_1^2x_2^2&-\frac{1}{\alpha_3^i}x_3&(\alpha_2^i)^2x_2^3+\left(-\frac{(\alpha_2^i)^4}{(\alpha_1^i)^2}+\frac{(\alpha_3^i)^2}{(\alpha_1^i)^2}\right)x_1^2x_2&-x_1^3
\end{array}\right)
\end{equation}
These manipulations leave the spectrum unchanged, except the condition $\langle\Omega\rangle\neq 0$. This matrix factorization is clearly reducible, since it contains two terms which are independent of $x_1,x_2$. Thus, we can make row-- and column manipulations to transform this factorization into a lower--dimensional one. A few steps of elementary operations yield the following simple result for $E_i^{red}$ (analogous steps lead to a corresponding expression for $J_i^{red}$): $E_i^{red}=\left(\begin{array}{cc}0&A_i\\B_i&0\end{array}\right)$, where
\begin{equation}
A_i=\left(\begin{array}{cc}
0&\frac{\alpha_1^i}{\alpha_2^i}\\
\frac{\alpha_2^i}{\alpha_1^i}\left(x_1^4+x_2^4-x_3^2+x_1^2x_2^2\left(-\frac{(\alpha_1^i)^2}{(\alpha_2^i)^2}-\frac{(\alpha_2^i)^2}{(\alpha_1^i)^2}+\frac{(\alpha_3^i)^2}{(\alpha_1^i)^2(\alpha_2^i)^2}\right)\right)&0
\end{array}\right)
\end{equation}
\begin{equation}
B_i=\left(\begin{array}{cc}\frac{1}{\alpha_3^1}x_3-\frac{1}{\alpha_1^i\alpha_2^i}x_1x_2&\frac{\alpha_1^i}{\alpha_2^i}x_1^2-\frac{\alpha_2^i}{\alpha_1^i}x_2^2\\
\frac{\alpha_2^i}{\alpha_1^i}x_1^2-\frac{\alpha_1^i}{\alpha_2^i}x_2^2&\alpha_3^ix_3+\frac{(\alpha_3^i)^2}{\alpha_1^i\alpha_2^i}x_1x_2\end{array}\right)
\end{equation}
Thus, the canonical $4\times 4$ matrix factorization for the long branes at a generic point in moduli space is isomorphic to the given $2\times 2$ factorization (\ref{long1}), (\ref{long2}).
\section{The Topological $A$--Model}
\label{acorr-sec}
\subsection{Computation of the Correlators}
To compare the results obtained in the $B$--model from the matrix factorizations,
we derive the $A$--model correlation functions here. The model, which is mirror to the Landau--Ginzburg orbifold we described in the previous sections, is a $T^2$ whose complex structure parameter is fixed to the value $e^{\frac{2\pi i}{4}}$.
In the $A$--model the correlators
can be obtained by an infinite sum over instanton areas\footnote{Since we are now in the A--model picture, $\tau$ is to be identified with the K\"ahler modulus.},
\begin{equation}
\langle \Phi_1...\Phi_N \rangle=\underset{N-gons}{\sum} e^{-2\pi i \tau A}.\label{eq:instsum}
\end{equation}
The (dimensionless) instanton areas are bounded by the branes on the compactified space.
Wilson lines contribute another factor in each term of Eq.~(\ref{eq:instsum}),
which equals the exponentiated integral around the boundary of the instanton.
The method was developed largely by
Polishchuk~\cite{Polishchuk:1998db,PolishchukFukaya,Polishchuk:2000kx,PolishchukTheta,PolishchukMirror}
and has since been applied once in a more physical setting~\cite{Govindarajan:2005im}.
Its application in phenomenology in~\cite{Ibanez:2001nd} has been very successful and the authors
were able to get the MSSM spectrum from it. In~\cite{Govindarajan:2005im}, 
a subset of factorizations, the 'long diagonal branes' as they were called
by the authors, were treated. To avoid a degenerate brane configuration
the branes can be placed a priori in such a way that no more than two branes intersect in
one point. By choosing appropriate boundary parameters $u_i$ (see below), arbitrary shifts
can be incorporated. 
\subsection{Three Point Functions for the Quartic Curve}
\begin{figure}
\begin{center}
\includegraphics{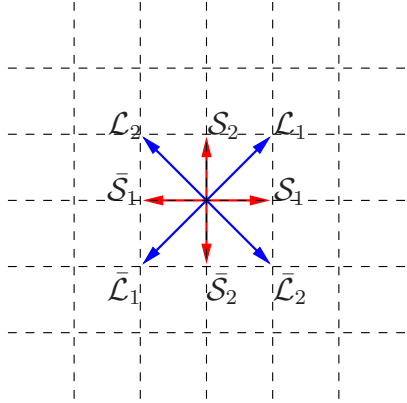}
\caption{Long and short branes on the covering space of the quartic torus.}\label{aquartic2}
\end{center}
\end{figure}
\begin{figure}[t]
\begin{center} 
\scalebox{1.0}{\includegraphics{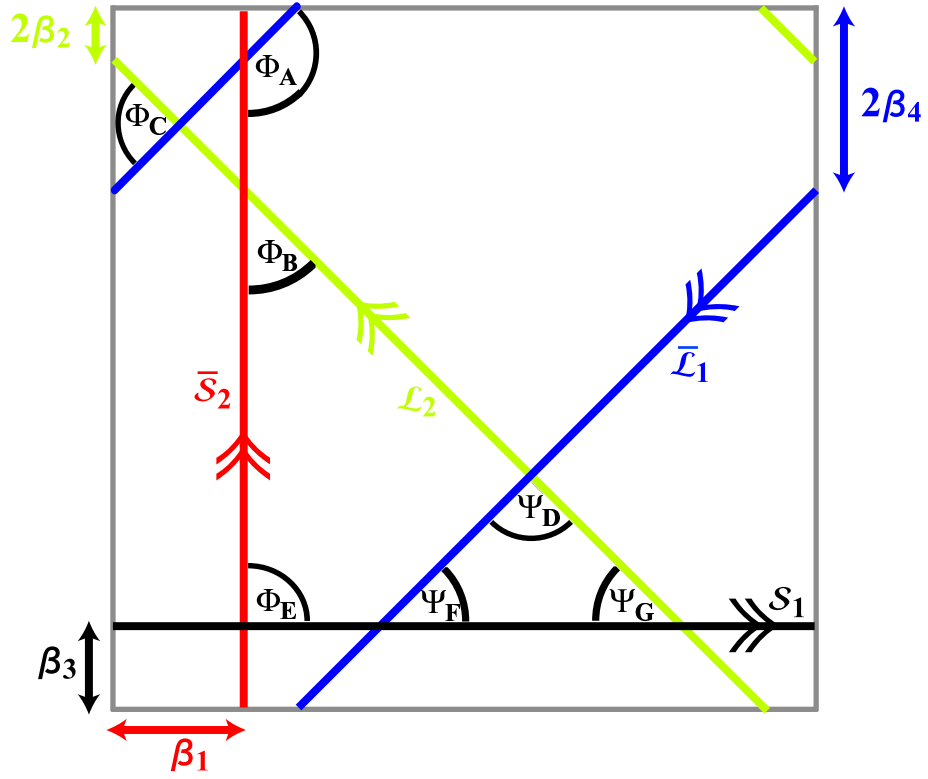}}
\caption{Covering space of the Quartic Torus.}\label{fig:covering}
\end{center}
\end{figure}
Fig. \ref{aquartic2} shows how the long and short branes wrap the torus.\\
Fig~\ref{fig:covering} shows the covering space of the quartic torus with
four branes on it. They are shifted by $\beta_i$
from the origin. By an appropriate choice of origin, two of the shifts can always be
set to zero if desired. The brane intersections labeled by $A$ through $G$ give rise to chiral
fields which are located at the positions,
\begin{eqnarray}
\begin{array}{ll}
A(\beta_1|1+\beta_1-2 \beta_4)& B(\beta_1|1-\beta_1-2 \beta_2)\\
C(-\beta_2+\beta_4|1-\beta_2-\beta_4)&
D(\frac{1}{2}-\beta_2 + \beta_4|\frac{1}{2}-\beta_2 - \beta_4)\\
E(\beta_1|\beta_3)& F(\beta_3+2 \beta_4|\beta_3)\\
G(1-2 \beta_2-\beta_3|\beta_3)
\end{array}
\end{eqnarray}
At each intersection one fermion or boson is shown. In the angles next
to it, the Serre dual field is located. The angles of the branes correspond to
their $R$--charges multiplied by $\pi$.
The two diagonal branes intersect each other twice in the fundamental domain, once
in $C$ and once in $D$. Consequently, for instantons bounded by two diagonal branes there
are always two choices of three--point functions.
\\
\begin{figure}[t]
\begin{center} 
\scalebox{0.67}{\includegraphics{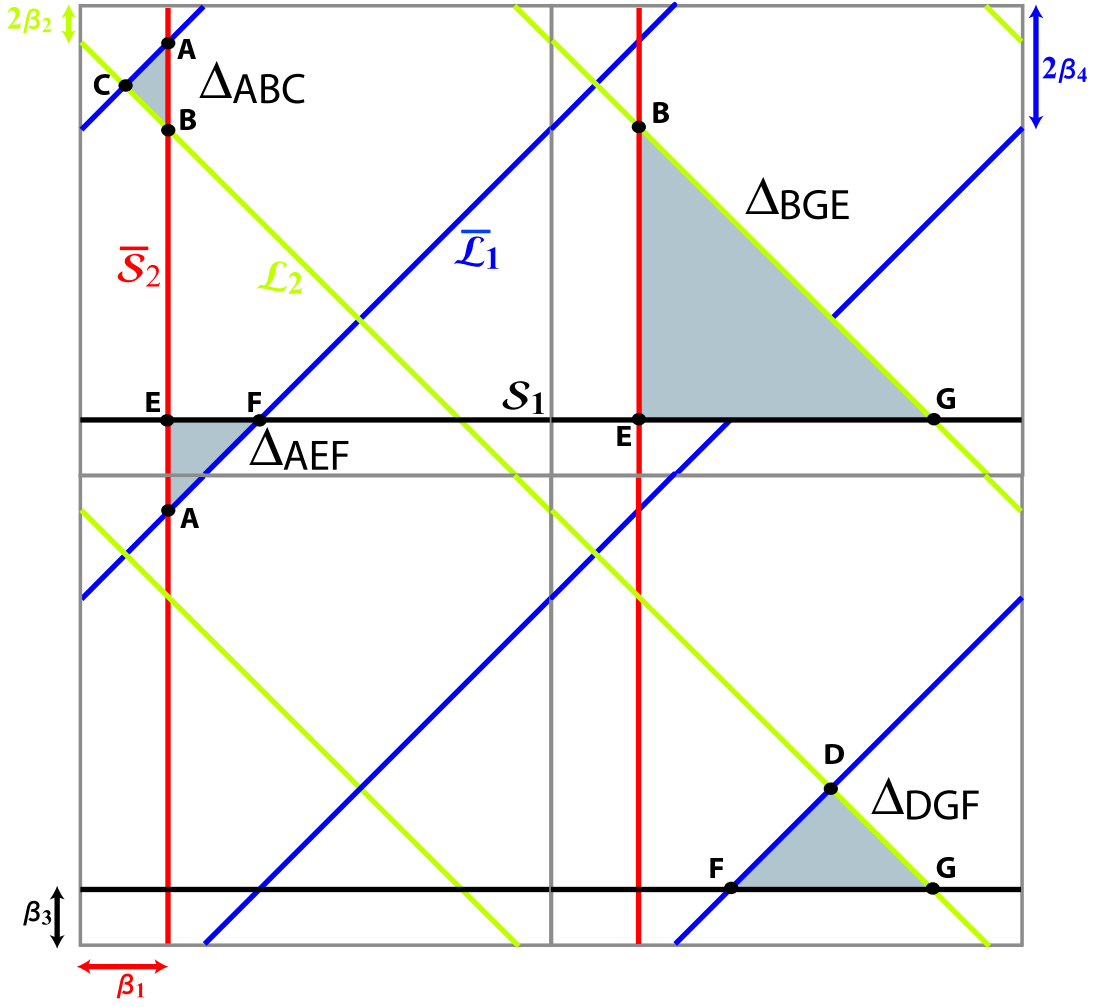}}
\caption{Three point functions on the torus.}\label{fig:3ptinstantons}
\end{center}
\end{figure}
The drawn brane configuration bounds altogether six different shapes of triangular instantons
which are the gray shaded areas in Fig.~\ref{fig:3ptinstantons}.
\begin{eqnarray}
\allowdisplaybreaks
\overline{BC} &=& \sqrt{2}(n+\beta_1 + \beta_2 - \beta_4)\nonumber\\
\overline{BD} &=& \sqrt{2}(n+\beta_1 + \beta_2 - \beta_4+1/2)\nonumber\\\\
A_{ABC}&=&\frac{1}{2}|\overline{BC}|^2=(n+\beta_1+\beta_2-\beta_4)^2\nonumber\\
A_{ABD}&=&\frac{1}{2}|\overline{BD}|^2=(n+\beta_1+\beta_2-\beta_4+1/2)^2,\nonumber\\
\nonumber\\
\overline{EF} &=& n-\beta_1+\beta_3 + 2\beta_4\nonumber\\
A_{AEF}&=&\frac{1}{2}|\overline{EF}|^2=\frac{1}{2}(n-\beta_1+\beta_3+2\beta_4)^2,\nonumber\\
\nonumber\\
\overline{CG} &=& \sqrt{2}(n+\beta_2 + \beta_3 + \beta_4)\nonumber\\
\overline{DG} &=& \sqrt{2}(n+\beta_2 + \beta_3 + \beta_4+1/2)\nonumber\\
A_{GFD}&=&\frac{1}{2}|\overline{DG}|^2=(n + \beta_2 + \beta_3 + \beta_4)^2\nonumber\\
A_{GFC}&=&\frac{1}{2}|\overline{DG}|^2=(n + \beta_2 + \beta_3 + \beta_4 + 1/2)^2,\nonumber\\
\nonumber\\
\overline{EG} &=& n+\beta_1+2\beta_2+\beta_3\nonumber\\
A_{BGE}&=&\frac{1}{2}|\overline{EG}|^2=\frac{1}{2}(n+\beta_1+2\beta_2+\beta_3)^2.
\end{eqnarray}
Expanding the squared brackets, the correlator $\langle ABC\rangle$ becomes
\begin{eqnarray}
\langle ABC \rangle \;\; \sim \;\; e^{2 \pi i \tau (\beta_1 + \beta_2 - \beta_4)^2}
\sum_{n=-\infty}^{\infty} e^{2 \pi i \tau [n^2+2n(\beta_1 + \beta_2 - \beta_4)]}\\
=e^{2 \pi i \tau (\beta_1 + \beta_2 - \beta_4)^2}
\sum_{n=-\infty}^{\infty} q^{n^2}e^{2\pi i n(2(\beta_1 + \beta_2 - \beta_4))\tau}.
\end{eqnarray}
Since we do not worry about the normalization here, we will suppress
the $\beta_i$--dependent prefactor in the following. If we include Wilson lines,
each summand must be multiplied with another factor depending on the lengths of the triangle sides,
\begin{equation}
e^{2 \pi i n(2(u_1^{\perp} + u_2^{\perp} - u_4^{\perp}))}.
\end{equation}
By defining $u_i=u_i^{\parallel}\tau + u_i^{\perp}$ where
$u_i^{\parallel}=\beta_i$ we can rewrite the correlators in the form
\begin{align}
\allowdisplaybreaks
\langle ABC \rangle&=\sum_{n=-\infty}^{\infty} q^{n^2}e^{2 \pi i n(2(u_1 + u_2 - u_4))}
=\Theta_3(2(u_1+u_2-u_4),2\tau),\\
\langle ABD \rangle&=\sum_{n=-\infty}^{\infty} q^{(n+1/2)^2}e^{2\pi i (n+1/2)(2(u_1 + u_2 - u_4))}
=\Theta_2(2(u_1+u_2-u_4),2\tau),\\
\langle AEF \rangle&=\sum_{n=-\infty}^{\infty} q^{\frac{1}{2}n^2}e^{2 \pi i n(-u_1 + u_3 + 2 u_4)}
=\Theta_3((-u_1+u_3+2u_4),\tau),\\
\langle GFD \rangle&=\sum_{n=-\infty}^{\infty} q^{n^2}e^{2 \pi i n(u_2 + u_3 + u_4)}
=\Theta_3(2(u_2+u_3+u_4),2\tau),\\
\langle GFC \rangle&=\sum_{n=-\infty}^{\infty} q^{(n+1/2)^2}e^{2\pi i (n+1/2)(2(u_2 + u_3 + u_4))}
=\Theta_2(2(u_2+u_3+u_4),2\tau),\\
\langle BGE \rangle&=\sum_{n=-\infty}^{\infty} q^{\frac{1}{2}n^2}e^{2 \pi i n(u_1 + 2 u_2 + u_3)}
=\Theta_3((u_1+2u_2+u_3),\tau).
\end{align}
The $B$--model correlators Eqs.~(\ref{eq:corr1})-(\ref{eq:corr4}) correspond to
the triangles $\langle ABC \rangle$, $\langle ABD \rangle$, $\langle GFC \rangle$ and
$\langle GFD \rangle$. Indeed the results are the same up to a small discrepancy:
Here we found $\Theta_2$ and $\Theta_3$ whereas in the $B$--model we found
$\Theta_1$ and $\Theta_4$ correlators. The former are theta
functions with characteristic $c_2=0$, the latter with $c_2=1/2$.
They are interrelated by a shift of the argument
\begin{eqnarray}
\Theta_2(u,\tau)&=&\Theta_1(u+1/2,\tau),\\
\Theta_3(u,\tau)&=&\Theta_4(u+1/2,\tau).
\end{eqnarray}
But this shift corresponds merely to a shift of origin on the covering space
of the torus and our results for the $A$-- and $B$--model are in complete
agreement.
\subsection{Higher Correlators}
\begin{figure}[!ht] 
\begin{center}
\scalebox{0.8}{\includegraphics{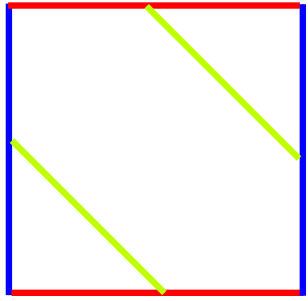}}\label{fig:quarticAside}
\caption{A configuration of three branes with
wrapping numbers $\mathcal{S}\simeq (0,1)$, $\mathcal{L}\simeq (1,-1)$, $\mathcal{S}\simeq (-1,0)$.}
\end{center}
\end{figure}

Correlators with more than three boundary insertions can be constructed
in the same manner as the three point functions. We will however
restrict ourselves to a configuration of three different branes.
If all brane shifts $\beta_i$ were zero, there is a degenerate intersection
in the origin. Consequently, the smallest $n$--gons have zero area
and can not be drawn on the covering space. In order to avoid confusion
we shifted the diagonal brane by $1/2$ but of course this shift could be
absorbed in the same manner in a prefactor as was done above with the
$\beta_i$. The configuration is displayed in Fig.~\ref{fig:quarticAside}.
In the following we proceed to determine the correlators for the shown configuration
of two short and one long brane.
The branes intersect only once on the covering space so the correlators
carry no label of the intersection number.\\
The three point correlator takes the form,
\begin{equation}
\begin{array}{rcl}
\Delta^{(i_3 i_1 i_2)}(\tau,u_i)
&=&\langle\Psi^{(i_3,i_1)}\Psi^{(i_1,i_2)}\Psi^{(i_2,i_3)}\rangle_{\mathrm{disk}}\\
&=&\Theta \left[\begin{array}{c} 1/2 \\ 0 \end{array}\right](\tau|u_1+u_2+u_3)\\
&=&{\displaystyle
\sum_{n=-\infty}^{\infty} q^{\frac{1}{2}(\frac{1}{2}+n)^2}
e^{2 \pi i (\frac{1}{2}+n)(u_1+u_2+u_3)}.}
\end{array}
\end{equation}
For the parallelogram correlator we get,
\begin{equation}
\begin{array}{rcl}
\mathcal{P}^{(i_4 i_1 i_2 i_3)}(\tau,u_i)
&=&\langle\Psi^{(i_4,i_1)}\Phi^{(i_1,i_2)}\Psi^{(i_2,i_3)}\Phi^{(i_3,i_4)}\rangle_{\mathrm{disk}}\\
&=&{\displaystyle -\sum_{m,n \in \mathbb{Z}}^{\mathrm{indef.}} q^{nm} e^{2 \pi i( n(u_1-u_3)+m(u_2-u_4))}}\\
&=&{\displaystyle -\sum_{m,n \in \mathbb{Z}-\{0\}}^{\mathrm{indef.}} q^{nm} e^{2 \pi i( n(u_1-u_3)+m(u_2-u_4))}}\\
& &{\displaystyle +\frac{1}{1-e^{2 \pi i (u_1-u_3)}}+\frac{1}{1-e^{2 \pi i (u_2-u_4)}}-1}
\end{array}
\end{equation}
where the indefinite sum is defined by,
\begin{equation}
\sum_{m,n \in \mathbb{Z}}^{\mathrm{indef.}} \equiv
\sum_{m,n=0}^{\infty}-\sum_{m,n=-\infty}^{-1}.
\end{equation}
The last equality in the definition of the parallelogram correlators made use of the
geometric series to sum up the terms where either $n$ or $m$ are zero. The result
is defined over a larger domain so this step is effectively an analytic continuation
of the expression. The same procedure can be used for the higher correlators below.\\
Note that the correlator carries an overall minus sign. That sign is missing in the
the parallelogram correlator in~\cite{Herbst:2006nn}. Without the sign, the expression
is not consistent with the $A_\infty$ relations. The sign swaps instantons
and anti-instantons and is a priori irrelevant, but must be consistent with the
choice of the other correlators.\\
The trapezoidal correlator was computed as,
\begin{equation}
\begin{array}{rcl}
\mathcal{T}^{(i_4 i_1 i_2 i_3)}(\tau,u_i)
&=&\langle\Psi^{(i_4,i_1)}\Psi^{(i_1,i_2)}\Phi^{(i_2,i_3)}\Phi^{(i_3,i_4)}\rangle_{\mathrm{disk}}\\
&=&{\displaystyle \sum_{m,n \in \mathbb{Z}}^{\mathrm{indef.}}  q^{\frac{1}{2}(2n+m+1)m}e^{2 \pi i (n+\frac{1}{2})(u_1-u_3) +m(u_1+u_2+u_4)}},
\end{array}
\end{equation}
where the indefinite sum is the same as defined above.\\
The five-point correlation function can be thought of as a parallelogram with one triangle subtracted and is given by,
\begin{equation}
\begin{array}{l}
\mathscr{P}^{(i_5 i_1 i_2 i_3 i_4)}(\tau,u_i)
=\langle\Psi^{(i_5,i_1)}\Phi^{(i_1,i_2)}\Phi^{(i_2,i_3)}\Phi^{(i_3,i_4)} \Phi^{(i_4,i_5)}\rangle_{\mathrm{disk}}\\
={\displaystyle \sum_{k,m,n \in \mathbb{Z}}^{\mathrm{indef.}}  q^{(m+k)(n+k)-\frac{1}{2}(\frac{1}{2}+k)^2}e^{2 \pi i\left( (n+k)(u_5-u_2) +(m+k)(u_1-u_4)+(k+\frac{1}{2})(u_3+u_2+u_4)\right)}}.
\end{array}
\end{equation}
Here, the indefinite sum with the three indices is defined by,
\begin{equation}
\sum_{k,m,n \in \mathbb{Z}}^{\mathrm{indef.}} \equiv
\sum_{k=0,\;m=1,\;n=1}^{\infty}-\sum_{k=-1,\;m=0\;,n=0}^{-\infty}.
\end{equation}
\\The hexagons in the instanton sum can be thought of as a parallelogram with the
two opposite acute-angled corners chopped off by subtracting two triangles.
This correlator completes the list of nonzero correlators and takes the form,
\begin{equation}
\begin{array}{l}
\mathcal{H}^{(i_6 i_1 i_2 i_3 i_4 i_5)}(\tau,u_i)
=\langle\Phi^{(i_6,i_1)}\Phi^{(i_1,i_2)}\Phi^{(i_2,i_3)}\Phi^{(i_3,i_4)} \Phi^{(i_4,i_5)} \Phi^{(i_5,i_6)}\rangle_{\mathrm{disk}}\\
={\displaystyle \sum_{k,l,m,n \in \mathbb{Z}}^{\mathrm{indef.}}  q^{m n-\frac{1}{2}(\frac{1}{2}+k)^2-\frac{1}{2}(\frac{1}{2}+l)^2}e^{2 \pi i\left( n(u_5-u_2) + m(u_1-u_4)+(k+\frac{1}{2})(u_3+u_2+u_4)+(l+\frac{1}{2})(-u_6-u_1-u_5)\right)}}.
\nonumber
\end{array}
\end{equation}
The sum runs as follows,
\begin{equation}
\sum_{k,m,n \in \mathbb{Z}}^{\mathrm{indef.}} \equiv
\sum_{n,m=0}^{\infty}\sum_{k,l=0}^{\mathrm{Min}(n,m)-1}
-\sum_{n,m=-1}^{-\infty} \sum_{\;\;\;k,l=\mathrm{Min}(n,m)-1}^{-1}.
\end{equation}
Since we computed topological disk amplitudes, the correlators should satisfy
the $\mathcal{A}_{\infty}$ relations~\cite{Herbst:2004jp}, which take the form
\begin{equation}
\displaystyle \sum_{k<l=1}^m (-1)^{\tilde a_1+...+\tilde a_k}
\mathcal F_{a_1...a_k c a_{l+1}...a_m} \mathcal{F}_{c a_{k+1}...a_l}=0\qquad m\ge 1.
\end{equation}
By an explicit computation we checked that the constraints are indeed satisfied
for all correlators.
\subsection{Short- vs. Long-Diagonal Branes and the Connection to the Cubic}
\begin{figure}[ht] 
\begin{center}
\scalebox{0.8}{\includegraphics{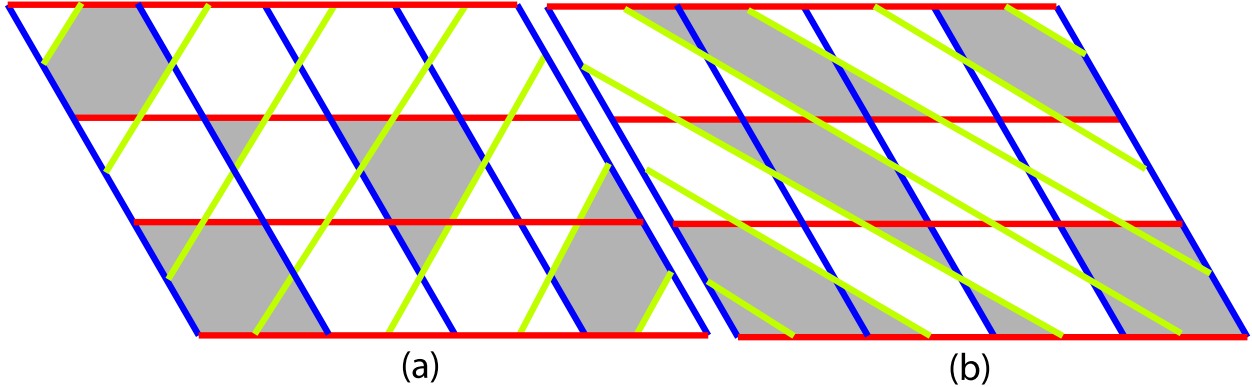}}\label{fig:cubicbranes}
\caption{A configuration with three short diagonal branes on the cubic is shown in
(a) and a configuration with two short--diagonal and one long--diagonal brane
is shown in (b).}
\end{center}
\end{figure}
In~\cite{Brunner:2004mt,Govindarajan:2005im,Herbst:2006nn}, the Cubic Curve
Landau--Ginzburg potential given by
\begin{equation}
W_{\mathbb{Z}_3}=x_1^3+x_2^3+x_3^3-a x_1 x_2 x_3,
\end{equation}
was discussed. This superpotential describes a $\mathbbm{Z}_3$ orbifold of $T^2$. Via mirror symmetry, this Landau--Ginzburg model corresponds in the $A$--model to a $T^2$ with complex structure parameter fixed to $e^{\frac{2\pi i}{3}}$. Since this model differs from the quartic torus only in the value of the complex structure modulus (respectively by the value of the K\"ahler modulus in the B--model) it must therefore
be possible to identify the amplitudes of the cubic with those of the quartic torus.
Indeed that can be done. The brane configuration corresponding to the one of the
Quartic torus in Fig.~\ref{fig:quarticAside} is shown in Fig.~\ref{fig:cubicbranes} for
the cubic. These correlators have not been computed in~\cite{Herbst:2006nn}, but
computing the instanton areas and setting up Eq.~(\ref{eq:instsum}) for this
case gives exactly the results derived the previous section for all correlators. 
It is interesting to note that the cubic curve configuration is described
by three short--diagonal branes, which can all be represented by the same $2\times 2$
matrix factorization Eq. (1.7) of~\cite{Govindarajan:2005im}. On the other hand,
for the quartic curve, two branes are the short branes
Eq.~(\ref{short1}) and (\ref{short2}) whereas the third is the long brane
Eq.~(\ref{long1}) and (\ref{long2}).
These branes not only look different in the matrix factorization framework,
they are also of different type in the CFT where they correspond to
the Recknagel-Schomerus~\cite{Recknagel:1997sb,Brunner:1999jq,Brunner:2005fv}
branes and the permutation branes~\cite{Recknagel:2002qq,Brunner:2005fv,Enger:2005jk}.
But even within the cubic description we find a dual description with one long
and two short branes, namely the branes with the wrapping numbers,
\begin{equation}
\mathcal{S}_1 \simeq (1,0),\;
\mathcal{S}_3 \simeq (0,1),\;
\mathcal{L}_1 \simeq (-1,-1),
\end{equation}
shown in Fig.~(\ref{fig:cubicbranes}). The triangles, parallelograms,
trapezoids as well as five-point and six-point shapes look different again,
but they have nevertheless same areas as those computed for the quartic torus
 and they give the same correlators. Symmetry considerations lead to the
conclusion that the following brane configurations are equivalent as well:
\begin{eqnarray}
\left\{\begin{array}{c}
(n_1,m_1)\\
(n_2,m_2)\\
(-n_1-n_2,-m_1-m_2)
\end{array}\right\}
\simeq
\left\{\begin{array}{c}
(n_1,-m_1)\\
(n_2,m_2)\\
(-n_1-n_2,m_1-m_2)
\end{array}\right\}\\
\nonumber\\
\left\{\begin{array}{c}
(n_1,m_1)\\
(n_2,m_2)\\
(-n_1-n_2,-m_1-m_2)
\end{array}\right\}
\simeq
\left\{ \begin{array}{c}
(-n_1,m_1)\\
(n_2,m_2)\\
(n_1-n_2,-m_1-m_2)
\end{array}\right\}.
\end{eqnarray}
In addition, there is a symmetry $n_i \longleftrightarrow m_i$.
This becomes clear when the configurations are visualized on the covering space
which the branes of these configurations slice into parts of equal areas.\\
\section{The Two--Variable Superpotential}
\label{twovarsec}
Matrix factorizations can be performed on models with both GSO--projections. The LG
superpotentials differ only by an additional variable, in our case $x_3$, which
enters the potential by a squared term, $x_3^2$. The following paragraphs deal with
the two--variable superpotential,
\begin{equation}
W_{LG}=x_1^4+x_2^4 - 2 \tilde a x_1^2 x_2^2.
\end{equation}
In this section we will consider this model with fixed boundary modulus $u$, keeping only the dependence of the complex structure modulus $\tau$ (via $\tilde{a}$). It turns out that, in this setup, the construction of permutation branes as given in \cite{Brunner:2005fv,Enger:2005jk} can be generalized in the presence of a bulk modulus. We show that it is posible to take over many results of the CFT calculations. 
\subsection{Rank $1$ Factorizations}
The basic factorizations of this potential turn out to be particularly
simple since the potential can be rewritten as,
\begin{equation}
W_{LG}=x_1^4+x_2^4 - 2 \tilde a x_1^2 x_2^2 = \overset{4}{\underset{n=1}{\displaystyle \prod}} (x_1-\eta_n x_2). 
\end{equation}
The four coefficients are,
\begin{equation}
\label{etadef}
\eta_{n}=\pm \sqrt{\tilde a \pm\sqrt{\tilde a^2-1}}\qquad n\in D=\{1,2,3,4\}.
\end{equation}
We use the convention $\eta_1\simeq (+,+),
\eta_2\simeq (+,-),\eta_3\simeq (-,+),\eta_4\simeq (-,-)$.
The variable transformation,
\begin{equation}
\tilde a \rightarrow \cosh(2 a),
\end{equation}
makes it possible to get rid of the square roots and brings
$\eta_n$ into the simple form,
\begin{equation}
\eta_n=\pm e^{\pm a}.
\end{equation}
We can now construct the rank 1 factorizations,
\begin{eqnarray}
Q_i^A=\begin{pmatrix}0&E^A\\J^A&0\end{pmatrix}\qquad
E_i^A=\underset{n \in D \backslash I_A}{\displaystyle \prod} (x_1-\eta_n x_2)\qquad
J_i^A=\underset{n \in I_A}{\displaystyle \prod} (x_1-\eta_n x_2). 
\end{eqnarray}
It remains to be checked that these branes satisfy the criterion of
orbifold invariance.
The $R$--matrix associated with the factorizations is defined by,
\begin{equation}
R_A=(2-\mbox{deg}(E_i^A))\mbox{diag}\left(\frac{1}{4},-\frac{1}{4}\right)
=(|I_A|-2)\mbox{diag}\left(\frac{1}{4},-\frac{1}{4}\right),
\end{equation}
with 'deg' denoting the polynomial degree and $|I_A|$ denoting the number of
elements in the index set. Related to $R_A$ are the orbifold matrices,
\begin{equation}
\gamma_i^A=\begin{pmatrix}1&0\\0&-1\end{pmatrix}e^{i \pi R^A} e^{-i \pi \varphi_i^A},\qquad (\gamma^A_i)^4=1,
\end{equation}
which define the orbifold action under which every $Q^i_A$
must be invariant,
\begin{equation}
\gamma_i^A Q_i^A(x_i e^{i \pi q_i})(\gamma_i^A)^{-1}=Q_i^A(x_i).
\end{equation}
A short computation shows that the listed factorizations are indeed stable
for four inequivalent
phases $\varphi_i$ of the factorizations. These phases are,
\begin{eqnarray}
\varphi_i=\left\{\begin{array}{l}
-\frac{1}{4}+i \frac{1}{2}\mbox{  for  }|I| \ne 2,\\
-\frac{1}{2}+i \frac{1}{2}\mbox{  for  }|I| = 2.
\end{array}\right.
\end{eqnarray}
The index $i$ labels the four orbifold copies.\\
We observe that despite the modulus $a$, the problem is very similar to the undeformed
theory, namely the tensor product of two identical $A$-series potentials,
$A_3 \otimes A_3$, which has already been studied~\cite{Ashok:2004zb,Brunner:2005fv}.
In many formulas the only difference is that the parameters $\eta_n$ of eq. (\ref{etadef}) 
appear at the place of the higher roots of $-1$ in the undeformed case. This will
enable us to take over many previous results with only slight modifications.\\
At this point we already note that the identification
of some matrix factorizations with the corresponding CFT description is
known for the tensor product models.
In particular, the Recknagel-Schomerus~\cite{Recknagel:1997sb,Brunner:1999jq,Brunner:2005fv}
branes and the permutation branes~\cite{Recknagel:2002qq,Brunner:2005fv,Enger:2005jk}
have been identified.
The equivalence between permutation branes and the minimal model is,
\begin{equation}
\|L,M,S_1=0,S_2=0 \rangle\rangle \Longleftrightarrow
\overset{(M+L)/2}{\underset{m=-(M+L)/2}{\displaystyle \prod}} (x_1-\eta_n x_2).
\end{equation}
\subsection{The Spectrum}
In order to avoid overloading the notation, from now on the orbifold
label $i$ will be suppressed.\\
The boundary preserving spectrum consists of no fermions, whereas
a basis of the bosons is given by the entire chiral ring 
$\mathbb{C}[x_1,x_2]/\mathcal{J}$, where $\mathcal{J}$ is the ideal associated
with $E^A$ and $J^A$.
More generally, for two different branes,
\begin{eqnarray}
E^A=\underset{n \in D \backslash I_A}{\displaystyle \prod} (x_1-\eta_n x_2)\qquad
J^A=\underset{n \in I_A}{\displaystyle \prod} (x_1-\eta_n x_2),\\
E^B=\underset{n \in D \backslash I_B}{\displaystyle \prod} (x_1-\eta_n x_2)\qquad
J^B=\underset{n \in I_B}{\displaystyle \prod} (x_1-\eta_n x_2), 
\end{eqnarray}
we find that the fermions $\Psi^{AB} \equiv \begin{pmatrix}0 & \psi^{AB}_0\\ \psi^{AB}_1 &0\end{pmatrix}$
which satisfy,
\begin{eqnarray}
\mathcal{D}(\Psi^{AB})\equiv Q^B\Psi^{AB}+\Psi^{AB} Q^A=0,
\end{eqnarray}
take the form,
\begin{eqnarray}
\begin{array}{rcl}
\psi^{AB}_0&=&-b^{AB}(x_1,x_2)\underset{n \in D\backslash I_A \cup I_B}{\displaystyle \prod} (x_1-\eta_n x_2),\\
\psi^{AB}_1&=&b^{AB}(x_1,x_2)\underset{n \in I_A \cap I_B}{\displaystyle \prod} (x_1-\eta_n x_2). 
\end{array}
\end{eqnarray}
The polynomials $b^{AB}(x_1,x_2)$ can take any value as long as $\Psi^{AB}$ does not become
an exact state. We will chose a basis for $b^{AB}(x_1,x_2)$ in the form,
\begin{equation}
b^{AB}(x_1,x_2)=\underset{n \in I^b}{\displaystyle \prod} (x_1-\eta_n x_2),
\end{equation}
where $I^b$ is an appropriate index set.
The coefficients  $c^{AB}(x_1,x_2)$ of the bosonic case can be dealt with analogously.\\
Altogether we count
$|I_A \backslash \{I_A \cap I_B\}|\cdot|I_B \backslash \{I_A \cap I_B\}|$ fermions
(see~\cite{Brunner:2005fv}).\\
The bosons $\Phi^{AB}= \begin{pmatrix} \phi^{AB}_0 & 0\\ 0& \phi^{AB}_1\end{pmatrix}$ are given by,
\begin{eqnarray}
\begin{array}{rcl}
\phi^{AB}_0&=&c^{AB}(x_1,x_2)\underset{n \in I_A \backslash I_A \cap I_B}{\displaystyle \prod} (x_1-\eta_n x_2),\\
\phi^{AB}_1&=&c^{AB}(x_1,x_2)\underset{n \in I_B \backslash I_A \cap I_B}{\displaystyle \prod} (x_1-\eta_n x_2). 
\end{array}\label{eq:bosons}
\end{eqnarray}
Their number is $|I_A \cap I_B|\cdot|D \backslash \{I_A \cup I_B\}|$.\\
\begin{figure}[t]
\begin{center} 
\scalebox{0.67}{\includegraphics{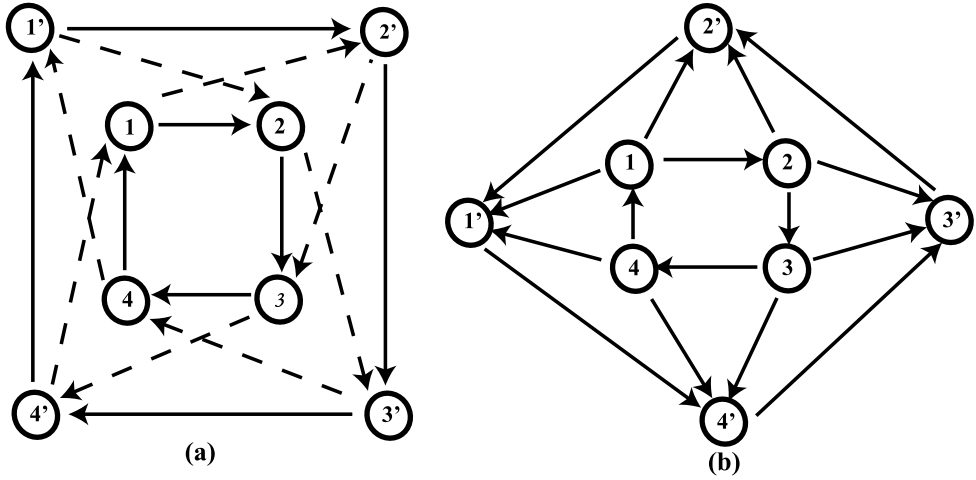}}\label{fig:quarticquiver}
\caption{The quiver diagrams for the two--variable quartic curve. Bosons are represented by solid lines, fermions by dashed lines.}
\end{center}
\end{figure}
The $R$--charge argument explained at the beginning of this work
gives the allowed charges for morphism between brane $Q$ and $Q'$,
\begin{eqnarray}
\begin{array}{rclcl}
ch(\phi)&=& \frac{1}{4}||I'|-|I|| \;\;(+\frac{1}{2})&\mbox{with}&
0 < ch(\phi) < 1,\\
ch(\psi)&=& \frac{1}{4}||I'|+|I||&\mbox{with}&
0 < ch(\psi) < 1.
\end{array}\label{eq:morcharge}
\end{eqnarray}
With these results it is possible to derive the quiver diagram.
Two different types of quivers exist; they are shown in Fig~\ref{fig:quarticquiver}.
Quiver (a) represents two branes of the type $|I|=|I'|=1$ where $I \ne I'$.
When a brane with $|I|\ne 2$ and one with $|I'|=2$ is present, the quiver is
the one in diagram (b). Depending on the index sets chosen, some brane labels
in the quiver can change and some bosons must be replaced by fermions.
Strings stretching between the primed and unprimed branes have charges
$1/4$ or $3/4$, the others carry charge $1/2$.
\subsection{The Three-Point Correlators for the Two--Variable Potential}
In the B--model, the topological three point disk
correlators of an $n$--variable theory can be computed directly from the generalized
residue~\cite{Kapustin:2003ga,Herbst:2004ax},
\begin{equation}
\langle \Phi_a^{AB}\Phi_b^{BC}\Phi_c^{CA}\rangle=\frac{1}{(2\pi i)^n}\oint\mathrm{d}^nx\frac{\mathrm{STr}\left((\partial Q^A)^{\wedge n}\Phi_a^{AB}\Phi_b^{BC}\Phi_c^{CA}\right)}{\partial_1 W\ldots\partial_n W}.
\label{eq:3ptfct}
\end{equation}
The selection rule for the 'suspended degree' tells us that the total
suspended $\mathbb{Z}_2$ degree in the two-variable case is even, i.e. we
must have either two fermionic insertions or none at all for the
correlator to be non-vanishing.
Another helpful selection rule is the restriction of the total $R$--charge
of the boundary insertions to be equal to the background charge,
\begin{equation}
\sum_i ch(\Phi_i)=1,\label{eq:bkch}
\end{equation}
which is $\hat c=1$ in the case of the torus. Taking a look at the morphism charges
Eq.~(\ref{eq:morcharge}), we see that the only way to satisfy
Eq.~(\ref{eq:bkch}) is by inserting a single operator
with charge $\frac{1}{2}$ and two with charge $\frac{1}{4}$.
Since the minimum charge of a fermion is $\frac{1}{2}$,
whereas the selection rule for the suspended degree does not allow just a single
fermionic insertion, we conclude that all three point functions with fermionic insertions
vanish and only bosonic insertions have to be considered. 
But this is inconsistent with the orbifold condition Eq.~(\ref{eq:orbi}).
A composition of morphisms mapping $A \rightarrow B \rightarrow C \rightarrow A$
gives us three versions of Eq.~(\ref{eq:orbi}) to be satisfied,
\begin{eqnarray}
ch(\Phi_{AB})=\varphi_B-\varphi_A+|\Phi_{AB}| \mbox{ mod 2},\\
ch(\Phi_{BC})=\varphi_C-\varphi_B+|\Phi_{BC}| \mbox{ mod 2},\\
ch(\Phi_{CA})=\varphi_A-\varphi_C+|\Phi_{CA}| \mbox{ mod 2}.
\end{eqnarray}
Adding them and comparing with Eq.~(\ref{eq:bkch}) we find,
\begin{equation}
|\Phi_{AB}|+|\Phi_{BC}|+|\Phi_{CA}|=1  \mbox{ mod 2}.
\end{equation}
That is, either one or all three morphism must be of odd degree, inconsistent
with the requirement that all be fermionic. \\
All bosonic matrices commute with each other since they are represented by diagonal
matrices, therefore the ordinary cyclic symmetry of the correlators
is elevated to an invariance under general permutations.\\
Using Eq.~(\ref{eq:bosons}) one gets,
\begin{eqnarray}
\phi^{AB}\phi^{BC}\phi^{CA}=\underset{i \in I_{ABC}}{\displaystyle \prod} (x_1-\eta_i x_2)\mathbbm{1}_{2\times 2},\\
I_{ABC}\equiv(I_A \cup I_B \cup I_C) \backslash (I_A \cap I_B \cap I_C).
\end{eqnarray}
From Eq.~(\ref{eq:bkch}) we also know that for nonzero correlators
$|I_{ABC}|=2$ must hold. For convenience, we introduce the notation,
\begin{equation}
\langle I_A, I_{ABC}\rangle\equiv \langle \phi^{AB}\phi^{BC}\phi^{CA} \rangle,
\end{equation}
for the correlators.
We are now ready to compute the actual values from Eq.~(\ref{eq:3ptfct}).
For $|I_A|=1$, we find,
\begin{eqnarray}
\begin{array}{l}
\displaystyle
\langle 2; 1,2 \rangle=-\langle 3; 1,2 \rangle=
\langle 4; 2,3 \rangle=-\langle 3; 2,3 \rangle=\frac{1}{\eta_1+\eta_4},\\\\
\displaystyle
\langle 1; 3,4 \rangle=-\langle 4; 3,4 \rangle=
\langle 1; 4,1 \rangle=-\langle 2; 4,1 \rangle=\frac{1}{\eta_1+\eta_4},\\\\
\displaystyle
\langle 1; 1,3 \rangle=-\langle 3; 1,3 \rangle=\frac{1}{2\eta_4},\\\\
\displaystyle
\langle 2; 2,4 \rangle=-\langle 4; 2,4 \rangle=\frac{1}{2\eta_1},
\end{array}
\end{eqnarray}
and for $|I_A|=2$ the result is,
\begin{eqnarray}
\begin{array}{l}
\displaystyle
\langle 1,2; 1,2 \rangle=-\langle 3,4; 1,2 \rangle=\langle 1,2; 3,4 \rangle=-\langle 3,4; 3,4 \rangle=\frac{1}{\eta_1+\eta_4},\\
\\
\displaystyle
\langle 1,2; 1,3 \rangle=-\langle 3,4; 1,3 \rangle=\langle 4,1; 1,3 \rangle=-\langle 2,3; 1,3 \rangle=\frac{1}{2\eta_4},\\
\\
\displaystyle
\langle 2,3; 2,3 \rangle=-\langle 4,1; 2,3 \rangle=\langle 2,3; 1,4 \rangle=-\langle 4,1; 1,4 \rangle=\frac{1}{\eta_1+\eta_2},\\
\\
\displaystyle
\langle 1,2; 2,4 \rangle=-\langle 4,1; 2,4 \rangle=\langle 2,3; 2,4 \rangle=-\langle 3,4; 2,4 \rangle=\frac{1}{2\eta_1}.
\end{array}
\end{eqnarray}
We observe that $\langle D\backslash I_A; I_{ABC} \rangle=-\langle I_A; I_{ABC} \rangle$,
so the case $|I_A|=3$ does not have to be treated separately. All correlators not listed vanish.
Since all three-point functions are of the same structure $1/(\eta_i+\eta_j)$,
the mapping is purely combinatorial.
\subsection{Higher Rank Factorizations}
Constructing $2 \times 2$ factorizations is straightforward, but 
the question is, whether they describe new branes
or how they can be obtained by tachyon condensation from rank 1 factorizations. 
In~\cite{Brunner:2005fv}, a clever transformation was found which relates
rank 1 factorizations of the type $x_1^d+x_2^d$ through tachyon
condensation to tensor product factorization. It will be shown here at
the example of the two--variable quartic curve that the transformation can be generalized to
apply to deformed potentials as well. We consider the following two branes,
\begin{eqnarray}
J_{P}=(x_1-\eta_n x_2)\qquad
E_{P}=\underset{m \ne n}{\displaystyle \prod} (x_1-\eta_i x_2),\\
J_{O}=\underset{m \ne n}{\displaystyle \prod} (x_1-\eta_i x_2)\qquad
E_{O}=(x_1-\eta_n x_2),
\end{eqnarray}
with $I_P \cap I_O=\emptyset$.
We define a BRST operator as follows:
\begin{equation}
Q=\begin{pmatrix}
0 & 0 & J_P & 0\\
0 & 0 & \lambda & J_O\\
E_P&0&0&0\\
-\lambda&E_O&0&0
\end{pmatrix}\mbox{ where }\lambda=x_1+\eta_n x_2.
\end{equation}
It corresponds to a new factorization obtained by tachyon condensation from
the two lower dimensional ones. (This new $Q$ is the so-called 'mapping cone'
of the 'cone construction'.)
It was shown that for the undeformed potential $a\rightarrow 0$, where
$\eta_i$ are fourth roots of $-1$, the above operator is isomorphic to,
\begin{equation}
\tilde Q=\begin{pmatrix}0&\tilde E\\ \tilde J&0\end{pmatrix},\qquad
\tilde E=\begin{pmatrix}x_2& x_1^3-2 a x_1 x_2^2\\ x_1&-x_2^3\end{pmatrix},
\tilde J=\begin{pmatrix}x_2^3& x_1^3-2 a x_1 x_2^2\\ x_1&-x_2\end{pmatrix}.
\label{eq:tens}
\end{equation}
The term with $a$, which vanishes in this limit, is kept here for later convenience.
The appropriate similarity transformation $\tilde Q=U^{-1}QU$ is given by,
\begin{equation}
U=\begin{pmatrix}
-\eta_n & 1 & 0 & 0\\
\eta_n & 1 & 0& 0\\
0& 0& 1 &-v\\
0&0& 0 &2 \eta_n
\end{pmatrix},
\qquad
v=\frac{x_2^3+\eta_n x_1^3}{x_1-\eta_n x_2} \mbox{ for }a=0.
\end{equation}
In order to include the deformations, it suffices to modify $v$ to,
\begin{equation}
v=\frac{x_2^3+\eta_n x_1^3-2a\eta_n x_1 x_2^2}{x_1-\eta_n x_2}.
\end{equation}
Now the limit $a\rightarrow 0$ can be relaxed.
It is important to note that the denominator in $v$ divides the nominator
without rest, this ensures that $U$ is indeed
an invertible matrix with polynomial entries. Note also that
the argument is not restrained to the torus potential. By adjusting
the exponential powers in the nominator of $v$, the argument holds for other
two variable $LG$--potentials with an appropriate deformation term as well,
provided that both fields have the same charge.\\
A short calculation shows that the obtained brane satisfies the orbifold condition,
and that again four copies of the branes with different phases of the factorization
exist.

\section{Conclusions}
In this paper we studied matrix factorizations and mirror symmetry on $T^2$, focusing, in the B--model, on the Landau--Ginzburg description related to the orbifold $T^2/\mathbbm{Z}_4$. In the B--model we identified the basic set of matrix factorizations of the Landau--Ginzburg superpotential. These factorizations correspond to long and short branes in the A--model picture. We computed the complete spectrum and calculated three--point functions. We could express the natural B--model variables $\alpha_i$ in terms of the geometric variables of the A--model by expressing them in terms of theta functions. This amounts to computing the mirror map. We could then check that our results are in agreement with the instanton sums in the A--model side, thus verifying homological mirror symmetry. \\
We also discussed the two--variable superpotential $x_1^4+x_2^4-2\tilde{a}x_1^2x_2^2$. For fixed boundary modulus, we were able to extend the construction of permutation branes to include the complex structure modulus of the torus.\\
As usual, there remain some open questions which give directions for future research. In this paper we gave some examples of A--model correlators with more than three insertions. For general Calabi--Yaus, it will be very hard to compute the instanton sums directly. For the simple case of the torus this is possible since the instantons are just the areas enclosed by the cycles winding around torus. For more complicated Calabi--Yaus one computes the quantum corrected amplitudes in the A--model by calculating the classical amplitudes in the B--model and relates them via mirror symmetry. However, so far, it is not known how to calculate $n$--point functions in the B--model. This may be achieved by CFT methods as presented in \cite{Herbst:2004jp} or by deformation theory methods introduced in \cite{Knapp:2006rd}. In either case, we expect that the presence of moduli will complicate the task significantly.\\
A further, related issue concerns the normalization of the states in the B--model. In this paper, we were rather careless about this problem. Our examples were so simple that we could, using the A--model result as a guideline, extract the correct values of the three--point functions without knowing how to correctly normalize the states.\\
Another interesting question is the extension of these methods to $T^6$ orbifolds. Toroidal compactifications are of great phenomenological interest. It is therefore suggestive to use the advantages of the matrix factorization description of D--branes to approach toroidal compactifications. Explicit results for the $T^2$ orbifolds may provide a first step in this direction.\\\\
{\bf Acknowledgments:} We want to thank Wolfgang Lerche for suggesting the topic of this paper. We are very grateful for his support and many interesting discussions. We also thank Hans Jockers for help and discussions. JK wants to thank Emanuel Scheidegger for sharing his knowledge on modular forms.
\appendix

\section{Cohomology}
We now state the explicit results for the boundary changing operators of the quartic torus.\\
Fermions and bosons will be denoted by:
\begin{equation}
\psi_{\mathcal{A}\mathcal{B}}(i,j)=\left(\begin{array}{cc}0&\psi^{(0)}_{\mathcal{A}\mathcal{B}}(i,j)\\
\psi^{(1)}_{\mathcal{A}\mathcal{B}}(i,j)&0
\end{array}\right)\quad\textrm{and}\quad
\phi_{\mathcal{A}\mathcal{B}}(i,j)=\left(\begin{array}{cc}\phi^{(0)}_{\mathcal{A}\mathcal{B}}(i,j)&0\\
0&\psi^{(1)}_{\mathcal{A}\mathcal{B}}(i,j)
\end{array}\right),
\end{equation}
respectively, where $\mathcal{A},\mathcal{B}$ denote the branes and $i,j$ denote the indices of the boundary moduli. We will use bars and tildes to make a distinction if there is more than one boundary state going from $\mathcal{A}$ to $\mathcal{B}$. Note that all states are defined only up to a normalization by a moduli--dependent factor.\\
\subsection{Long Branes}
From Fig. 1 one can read off that there are two fermions and two bosons stretching between two of the long branes. All of them have R--charge $1/2$. The first fermion has the following structure:
\begin{equation}
\psi^{(0)}_{\mathcal{L}_1\mathcal{L}_2}(1,2)=\left(\begin{array}{cc}a(1,2)&b(1,2)\\
c(1,2)&d(1,2)
\end{array}\right)\quad
\psi^{(1)}_{\mathcal{L}_1\mathcal{L}_2}(1,2)=\left(\begin{array}{cc}
d(2,1)&-b(2,1)\\
-c(2,1)&a(2,1)
\end{array}\right),
\end{equation}
where
\begin{equation*}
a(1,2)=\left(\frac{\alpha_2^2}{\alpha_1^1}+\frac{\alpha_2^1\alpha_3^2}{\alpha_1^2\alpha_3^1}\right)x_2\qquad b(1,2)=\left(-\frac{(\alpha_1^1)^2\alpha_1^2}{\alpha_2^1\alpha_3^1\alpha_3^2}+\frac{\alpha_2^1(\alpha_2^2)^2}{\alpha_1^2\alpha_3^1\alpha_3^2}\right)x_1
\end{equation*}
\begin{equation}
c(1,2)=\left(\frac{\alpha_1^1(\alpha_1^2)^2}{\alpha_2^2}-\frac{(\alpha_2^1)^2\alpha_2^2}{\alpha_1^1}\right)x_1\qquad d(1,2)=\left(\frac{-\alpha_1^1}{\alpha_2^2}-\frac{\alpha_1^2\alpha_3^1}{\alpha_2^1\alpha_3^2}\right)x_2
\end{equation}
The second fermion has the same structure with $x_1$ and $x_2$ exchanged:
\begin{equation}
\bar{\psi}^{(0)}_{\mathcal{L}_1\mathcal{L}_2}(1,2)=\left(\begin{array}{cc}\bar{a}(1,2)&\bar{b}(1,2)\\
\bar{c}(1,2)&\bar{d}(1,2)
\end{array}\right)\quad
\bar{\psi}^{(1)}_{\mathcal{L}_1\mathcal{L}_2}(1,2)=\left(\begin{array}{cc}
\bar{d}(2,1)&-\bar{b}(2,1)\\
-\bar{c}(2,1)&\bar{a}(2,1)
\end{array}\right),
\end{equation}
where
\begin{equation*}
\bar{a}(1,2)=\left(\frac{1}{\alpha_1^1\alpha_2^1}+\frac{\alpha_3^2}{\alpha_1^2\alpha_2^2\alpha_3^1}\right)x_1\qquad \bar{b}(1,2)=\left(-\frac{\alpha_1^2}{\alpha_2^2\alpha_3^1\alpha_3^2}+\frac{(\alpha_2^1)^2\alpha_2^2}{(\alpha_1^1)^2\alpha_1^2\alpha_3^1\alpha_3^2}\right)x_2
\end{equation*}
\begin{equation}
\bar{c}(1,2)=\left(\frac{\alpha_1^1}{\alpha_2^1}-\frac{\alpha_2^1(\alpha_2^2)^2}{\alpha_1^1(\alpha_1^2)^2}\right)x_2\qquad\bar{d}(1,2)=\left(-\frac{\alpha_2^1}{\alpha_1^1(\alpha_1^2)^2}-\frac{\alpha_2^2\alpha_3^1}{(\alpha_1^1)^2\alpha_1^2\alpha_3^2}\right)x_1
\end{equation}
The first boson looks as follows:
\begin{equation}
\phi^{(0)}_{\mathcal{L}_1\mathcal{L}_2}(1,2)=\left(\begin{array}{cc}a(1,2)&b(1,2)\\
c(1,2)&d(1,2)
\end{array}\right)\quad
\phi^{(1)}_{\mathcal{L}_1\mathcal{L}_2}(1,2)=\left(\begin{array}{cc}
\frac{\alpha_3^2}{\alpha_3^1}d(1,2)&\frac{1}{\alpha_3^1\alpha_3^2}c(1,2)\\
\alpha_3^1\alpha_3^2 b(1,2)&\frac{\alpha_3^1}{\alpha_3^2}a(1,2)
\end{array}\right),
\end{equation}
where
\begin{equation*}
a(1,2)=\left(\alpha_1^1+\frac{(\alpha_1^1)^2\alpha_2^1\alpha_3^2}{\alpha_1^2\alpha_2^2\alpha_3^1}\right)x_2\qquad b(1,2)=\left(-\frac{(\alpha_1^1)^2\alpha_1^2\alpha_2^1}{\alpha_2^2\alpha_3^1\alpha_3^2}+\frac{(\alpha_1^1)^4\alpha_2^2}{\alpha_1^2\alpha_2^1\alpha_3^1\alpha_3^2}\right)x_1
\end{equation*}
\begin{equation}
c(1,2)=\left(-\alpha_1^1(\alpha_2^1)^2+\frac{(\alpha_1^1)^3(\alpha_2^2)^2}{(\alpha_1^2)^2}\right)x_1\qquad d(1,2)=\left(\frac{(\alpha_1^1)^3}{(\alpha_1^2)^2}+\frac{(\alpha_1^1)^2\alpha_2^2\alpha_3^1}{\alpha_1^2\alpha_2^1\alpha_3^2}\right)x_2
\end{equation}
For the second boson we find:
\begin{equation}
\bar{\phi}^{(0)}_{\mathcal{L}_1\mathcal{L}_2}(1,2)=\left(\begin{array}{cc}\bar{a}(1,2)&\bar{b}(1,2)\\
\bar{c}(1,2)&\bar{d}(1,2)
\end{array}\right)\quad
\bar{\phi}^{(1)}_{\mathcal{L}_1\mathcal{L}_2}(1,2)\left(\begin{array}{cc}
\frac{\alpha_3^1}{\alpha_3^2}d(1,2)&\frac{1}{\alpha_3^1\alpha_3^2}c(1,2)\\
\alpha_3^1\alpha_3^2 b(1,2)&\frac{\alpha_3^1}{\alpha_3^2}a(1,2)
\end{array}\right),
\end{equation}
where
\begin{equation*}
\bar{a}(1,2)=\left(\frac{\alpha_1^1}{\alpha_1^2}+\frac{\alpha_2^2\alpha_3^1}{\alpha_2^1\alpha_3^2}\right)x_1\qquad\bar{b}(1,2)=\left(-\frac{\alpha_1^2(\alpha_2^1)^2}{\alpha_1^1(\alpha_3^2)^2}+\frac{\alpha_1^1(\alpha_2^2)^2}{\alpha_1^2(\alpha_3^2)^2}\right)x_2
\end{equation*}
\begin{equation}
\bar{c}(1,2)=\left(-\frac{(\alpha_1^2)^2\alpha_2^1\alpha_3^1}{\alpha_2^2\alpha_3^2}+\frac{(\alpha_1^1)^2\alpha_2^2\alpha_3^1}{\alpha_2^1\alpha_3^2}\right)x_2\qquad\bar{d}(1,2)=\left(\frac{\alpha_1^2(\alpha_3^1)^2}{\alpha_1^1(\alpha_3^2)^2}+\frac{\alpha_2^1\alpha_3^1}{\alpha_2^2\alpha_3^2}\right)x_1
\end{equation}

\subsection{Short Branes}
There is one charge $1/2$ fermion and one charge $1/2$ boson. The fermion is given by:
\begin{equation}
\psi^{(0)}_{\mathcal{S}_1\mathcal{S}_2}=\left(\begin{array}{cc}
a(1,2)&b(1,2)\\
c(1,2)&d(1,2)
\end{array}\right)\qquad
\psi^{(1)}_{\mathcal{S}_1\mathcal{S}_2}=\left(\begin{array}{cc}
-\frac{\alpha_3^2}{\alpha_3^1}d(1,2)&\alpha_3^1\alpha_3^2c(1,2)\\
\frac{1}{\alpha_3^1\alpha_3^2}b(1,2)&-\frac{\alpha_3^2}{\alpha_3^1}a(1,2)
\end{array}\right),
\end{equation}
where
\begin{eqnarray}
a(1,2)&=&\frac{\alpha_1^2\alpha_2^1\alpha_2^2}{\alpha_1^1}+\frac{(\alpha_2^1)^2\alpha_3^2}{\alpha_3^1}\nonumber \\
b(1,2)&=&\left(\frac{\alpha_1^1\alpha_1^2\alpha_2^1\alpha_3^2}{\alpha_3^1}+\frac{(\alpha_1^2)^2(\alpha_2^1)^2\alpha_3^2}{\alpha_2^2\alpha_3^1}\right)x_2+\left(\frac{(\alpha_2^1)^3\alpha_2^2\alpha_3^2}{\alpha_1^1\alpha_3^1}+\frac{(\alpha_2^1)^2(\alpha_2^2)^2\alpha_3^2}{\alpha_1^2\alpha_3^1}\right)x_1\nonumber\\
c(1,2)&=&\left(\frac{(\alpha_1^2)^2\alpha_2^1}{(\alpha_3^1)^2}+\frac{\alpha_1^1\alpha_1^2\alpha_2^2}{(\alpha_3^1)^2}\right)x_2+\left(\frac{\alpha_1^2(\alpha_2^1)^3\alpha_2^2}{(\alpha_1^1)^2(\alpha_3^1)^2}+\frac{(\alpha_2^1)^2(\alpha_2^2)^2}{\alpha_1^1(\alpha_3^1)^2}\right)x_1\nonumber \\
d(1,2)&=&\left(\frac{\alpha_2^1\alpha_2^2\alpha_3^2}{(\alpha_1^1)^2\alpha_3^1}+\frac{(\alpha_2^1)^2(\alpha_3^2)^2}{\alpha_1^1\alpha_1^2(\alpha_3^1)^2}\right)x_1^2+\left(\frac{\alpha_1^2\alpha_3^2}{\alpha_1^1\alpha_3^1}+\frac{\alpha_2^1(\alpha_3^2)^2}{\alpha_2^2(\alpha_3^1)^2}\right)x_2^2
\end{eqnarray}
The boson looks as follows:
\begin{equation}
\phi^{(0)}_{\mathcal{S}_1\mathcal{S}_2}=\left(\begin{array}{cc}
a(1,2)&b(1,2)\\
c(1,2)&d(1,2)
\end{array}\right)\qquad
\phi^{(1)}_{\mathcal{S}_1\mathcal{S}_2}=\left(\begin{array}{cc}
\frac{\alpha_3^1}{\alpha_3^2}d(1,2)&-\alpha_3^1\alpha_3^2c(1,2)\\
-\frac{1}{\alpha_3^1\alpha_3^2}b(1,2)&\frac{\alpha_3^2}{\alpha_3^1}a(1,2)
\end{array}\right),
\end{equation}
where
\begin{eqnarray}
a(1,2)&=&\left(\alpha_3^1+\frac{\alpha_1^2\alpha_2^1\alpha_3^1}{\alpha_1^1\alpha_2^2}\right)x_2+\left(\frac{(\alpha_2^1)^2\alpha_2^2\alpha_3^1}{(\alpha_1^1)^2\alpha_1^2}+\frac{\alpha_2^1(\alpha_2^2)^2\alpha_3^1}{\alpha_1^1(\alpha_1^2)^2}\right)x_1\nonumber \\
b(1,2)&=&\frac{\alpha_2^2(\alpha_3^1)^2\alpha_3^2}{(\alpha_1^1)^2}-\frac{\alpha_2^1\alpha_3^1(\alpha_3^2)^2}{\alpha_1^1\alpha_1^2}\nonumber\\
c(1,2)&=&\left(\frac{\alpha_2^2\alpha_3^1}{(\alpha_1^1)^3\alpha_1^2}-\frac{\alpha_2^1\alpha_3^2}{(\alpha_1^1)^2(\alpha_1^2)^2}\right)x_1^2+\left(\frac{\alpha_3^1}{(\alpha_1^1)^2\alpha_2^1}-\frac{\alpha_3^2}{\alpha_1^1\alpha_1^2\alpha_2^2}\right)x_2^2\nonumber\\
d(1,2)&=&\left(\frac{\alpha_1^2\alpha_3^2}{\alpha_1^1}+\frac{\alpha_2^2\alpha_3^2}{\alpha_2^1}\right)x_2+\left(\frac{(\alpha_2^1)^2\alpha_2^2\alpha_3^2}{(\alpha_1^1)^3}+\frac{\alpha_2^1(\alpha_2^2)^2\alpha_3^2}{(\alpha_1^1)^2\alpha_1^2}\right)x_1
\end{eqnarray}

\subsection{Short to Long}
There is one fermion with charge $1/4$ and one with charge $3/4$ and, symmetrically, bosons with charges $1/4$ and $3/4$. We will denote the $3/4$--states with tildes.
The charge $1/4$ fermion reads:
\begin{equation}
\psi^{(0)}_{\mathcal{S}_1\mathcal{L}_2}=\left(\begin{array}{cc}
a(1,2)&b(1,2)\\
c(1,2)&d(1,2)
\end{array}\right)\qquad
\psi^{(1)}_{\mathcal{S}_1\mathcal{L}_2}=\left(\begin{array}{cc}
-\alpha_3^1\alpha_3^2d(1,2)&-\frac{\alpha_3^1}{\alpha_3^2}c(1,2)\\
\frac{\alpha_3^2}{\alpha_3^1}b(1,2)&\frac{1}{\alpha_3^1\alpha_3^2}a(1,2)
\end{array}\right),
\end{equation}
where
\begin{eqnarray}
a(1,2)&=&\frac{\alpha_1^1\alpha_3^1}{(\alpha_2^1)^2}-\frac{(\alpha_1^1)^2\alpha_3^2}{\alpha_1^2\alpha_2^1\alpha_2^2}\nonumber \\
b(1,2)&=&-\frac{(\alpha_1^1)^3\alpha_1^2}{(\alpha_2^1)^2\alpha_2^2\alpha_3^2}+\frac{\alpha_1^1\alpha_2^2}{\alpha_1^2\alpha_3^2}\nonumber \\
c(1,2)&=&\left(\frac{1}{\alpha_3^1}-\frac{(\alpha_1^1)^2(\alpha_1^2)^2}{(\alpha_2^1)^2(\alpha_2^2)^2\alpha_3^1}\right)x_1+\left(\frac{(\alpha_1^1)^3}{(\alpha_2^1)^3\alpha_3^1}-\frac{\alpha_1^1(\alpha_2^2)^2}{(\alpha_1^2)^2\alpha_2^1\alpha_3^1}\right)x_2\nonumber\\
d(1,2)&=&\left(\frac{(\alpha_1^1)^2}{(\alpha_2^1)^2(\alpha_2^2)^2\alpha_3^1}-\frac{\alpha_1^1\alpha_1^2}{(\alpha_2^1)^3\alpha_2^2\alpha_3^2}\right)x_2+\left(-\frac{\alpha_1^1}{(\alpha_1^2)^2\alpha_2^1\alpha_3^1}+\frac{\alpha_2^2}{\alpha_1^2(\alpha_2^1)^2\alpha_3^2}\right)x_1
\end{eqnarray}
When computing the charge $3/4$ states one has to be careful with exact states. We can use those to ``gauge away'' one of the variables. We choose the convention that the charge $3/4$ states contain terms linear and quadratic in $x_1$ but not linear and quadratic in $x_2$. One mixed term $x_1x_2$ will always remain.
In this gauge the charge $3/4$ fermion reads:
\begin{equation}
\tilde{\psi}^{(0)}_{\mathcal{S}_1\mathcal{L}_2}=\left(\begin{array}{cc}
\tilde{a}(1,2)&\tilde{b}(1,2)\\
\tilde{c}(1,2)&\tilde{d}(1,2)
\end{array}\right)\qquad
\tilde{\psi}^{(1)}_{\mathcal{S}_1\mathcal{L}_2}=\left(\begin{array}{cc}
-\alpha_3^1\alpha_3^2\tilde{d}(1,2)&-\frac{\alpha_3^1}{\alpha_3^2}\tilde{c}(1,2)\\
\frac{\alpha_3^2}{\alpha_3^1}\tilde{b}(1,2)&\frac{1}{\alpha_3^1\alpha_3^2}\tilde{a}(1,2)
\end{array}\right),
\end{equation}
where
\begin{eqnarray}
\tilde{a}(1,2)&=&\left(-(\alpha_1^1)^2\alpha_3^2+\frac{(\alpha_1^2)^2(\alpha_2^1)^2\alpha_3^2}{(\alpha_2^2)^2}\right)x_1\nonumber\\
\tilde{b}(1,2)&=&\left(\frac{\alpha_1^2\alpha_3^1}{\alpha_2^2}+\frac{\alpha_1^1\alpha_2^1\alpha_3^2}{(\alpha_2^2)^2}\right)x_1\nonumber\\
\tilde{c}(1,2)&=&\left(\frac{(\alpha_1^2)^2\alpha_3^2}{\alpha_1^1(\alpha_2^2)^2}+\frac{\alpha_1^2\alpha_2^1(\alpha_3^2)^2}{(\alpha_2^2)^2\alpha_3^1}\right)x_1^2+\left(-\frac{\alpha_3^2}{\alpha_2^1}-\frac{\alpha_1^1(\alpha_3^2)^2}{\alpha_1^2\alpha_2^2\alpha_3^1}\right)x_1x_2\nonumber\\
\tilde{d}(1,2)&=&\left(-\frac{(\alpha_1^2)^3\alpha_2^1}{(\alpha_2^2)^3\alpha_3^1}+\frac{(\alpha_1^1)^2\alpha_1^2}{\alpha_2^1\alpha_2^2\alpha_3^1}\right)x_1x_2+\left(\frac{\alpha_1^2(\alpha_2^1)^2}{\alpha_1^1\alpha_2^2\alpha_3^1}-\frac{\alpha_1^1\alpha_2^2}{\alpha_1^2\alpha_3^1}\right)x_1^2
\end{eqnarray}
The charge $1/4$ boson looks as follows:
\begin{equation}
\phi^{(0)}_{\mathcal{S}_1\mathcal{L}_2}=\left(\begin{array}{cc}
a(1,2)&b(1,2)\\
c(1,2)&d(1,2)
\end{array}\right)\qquad
\phi^{(1)}_{\mathcal{S}_1\mathcal{L}_2}=\left(\begin{array}{cc}
-\alpha_3^1\alpha_3^2d(1,2)&-\frac{\alpha_3^1}{\alpha_3^2}c(1,2)\\
\frac{\alpha_3^2}{\alpha_3^1}b(1,2)&\frac{1}{\alpha_3^1\alpha_3^2}a(1,2)
\end{array}\right),
\end{equation}
where
\begin{eqnarray}
a(1,2)&=&\frac{\alpha_2^1\alpha_3^1}{(\alpha_1^1)^2}-\frac{(\alpha_2^1)^2\alpha_3^2}{\alpha_1^1\alpha_1^2\alpha_2^2}\nonumber\\
b(1,2)&=&-\frac{\alpha_1^2(\alpha_2^1)^2}{(\alpha_1^1)^2\alpha_2^2\alpha_3^2}+\frac{\alpha_2^1\alpha_2^2}{\alpha_1^2\alpha_3^2}\nonumber\\
c(1,2)&=&\left(\frac{1}{\alpha_3^1}-\frac{(\alpha_1^2)^2(\alpha_2^1)^2}{(\alpha_1^1)^2(\alpha_2^2)^2\alpha_3^1}\right)x_2+\left(\frac{(\alpha_2^1)^3}{(\alpha_1^1)^3\alpha_3^1}-\frac{\alpha_2^1(\alpha_2^2)^2}{\alpha_1^1(\alpha_1^2)^2\alpha_3^1}\right)x_1\nonumber\\
d(1,2)&=&\left(\frac{(\alpha_2^1)^2}{(\alpha_1^1)^2(\alpha_2^2)^2\alpha_3^1}-\frac{\alpha_1^2\alpha_2^1}{(\alpha_1^1)^3\alpha_2^2\alpha_3^2}\right)x_1+\left(-\frac{\alpha_2^1}{\alpha_1^1(\alpha_1^2)^2\alpha_3^1}+\frac{\alpha_2^2}{(\alpha_1^1)^2\alpha_1^2\alpha_3^2}\right)x_2
\end{eqnarray}
The charge $3/4$ boson is:
\begin{equation}
\tilde{\phi}^{(0)}_{\mathcal{S}_1\mathcal{L}_2}=\left(\begin{array}{cc}
\tilde{a}(1,2)&\tilde{b}(1,2)\\
\tilde{c}(1,2)&\tilde{d}(1,2)
\end{array}\right)\qquad
\tilde{\phi}^{(1)}_{\mathcal{S}_1\mathcal{L}_2}=\left(\begin{array}{cc}
-\alpha_3^1\alpha_3^2\tilde{d}(1,2)&-\frac{\alpha_3^1}{\alpha_3^2}\tilde{c}(1,2)\\
\frac{\alpha_3^2}{\alpha_3^1}\tilde{b}(1,2)&\frac{1}{\alpha_3^1\alpha_3^2}\tilde{a}(1,2)
\end{array}\right),
\end{equation}
where
\begin{eqnarray}
\tilde{a}(1,2)&=&\left(-\frac{\alpha_1^1\alpha_2^2\alpha_3^1}{\alpha_1^2}+\frac{(\alpha_1^1)^2\alpha_2^1\alpha_3^2}{(\alpha_1^2)^2}\right)x_1\nonumber\\
\tilde{b}(1,2)&=&\left(\frac{\alpha_1^1(\alpha_2^1)^2}{\alpha_3^2}-\frac{(\alpha_1^1)^3(\alpha_2^2)^2}{(\alpha_1^2)^2\alpha_3^2}\right)x_1\nonumber\\
\tilde{c}(1,2)&=&\left(\frac{\alpha_1^1\alpha_1^2\alpha_2^1}{\alpha_2^1\alpha_3^1}-\frac{(\alpha_1^1)^3\alpha_2^2}{\alpha_1^2\alpha_2^1\alpha_3^1}\right)x_1x_2+\left(-\frac{(\alpha_2^1)^2\alpha_2^2}{\alpha_1^2\alpha_3^1}+\frac{(\alpha_1^1)^2(\alpha_2^2)^3}{(\alpha_1^2)^3\alpha_3^1}\right)x_1^2\nonumber\\
\tilde{d}(1,2)&=&\left(-\frac{\alpha_1^1\alpha_2^1}{\alpha_1^2\alpha_2^2\alpha_3^1}+\frac{1}{\alpha_3^2}\right)x_1^2+\left(\frac{(\alpha_1^1)^2\alpha_2^2}{(\alpha_1^2)^3\alpha_3^1}-\frac{\alpha_1^1(\alpha_2^2)^2}{(\alpha_1^2)^2\alpha_2^1\alpha_3^2}\right)x_1x_2
\end{eqnarray}

\subsection{Long to Short}
By Serre duality, the bosons and fermions pair up with the states going from the short branes to the long branes.
The charge $1/4$ fermion reads:
\begin{equation}
\psi^{(0)}_{\mathcal{L}_1\mathcal{S}_2}=\left(\begin{array}{cc}
a(1,2)&b(1,2)\\
c(1,2)&d(1,2)
\end{array}\right)\qquad
\psi^{(1)}_{\mathcal{L}_1\mathcal{S}_2}=\left(\begin{array}{cc}
\frac{1}{\alpha_3^1\alpha_3^2}d(1,2)&-\frac{\alpha_3^2}{\alpha_3^1}c(1,2)\\
\frac{\alpha_3^1}{\alpha_3^2}b(1,2)&-{\alpha_3^1\alpha_3^2}a(1,2)
\end{array}\right),
\end{equation}
where
\begin{eqnarray}
a(1,2)&=&\frac{1}{\alpha_3^1}-\frac{\alpha_1^1\alpha_2^1\alpha_3^2}{\alpha_1^2\alpha_2^2(\alpha_3^1)^2}\nonumber\\
b(1,2)&=&\left(\frac{(\alpha_2^1)^3\alpha_3^2}{\alpha_1^1\alpha_1^2(\alpha_3^1)^2}-\frac{\alpha_1^1\alpha_1^2\alpha_2^1\alpha_3^2}{(\alpha_2^2)^2(\alpha_3^1)^2}\right)x_2+\left(\frac{(\alpha_1^1)^3\alpha_3^2}{\alpha_2^1\alpha_2^2(\alpha_3^1)^2}-\frac{\alpha_1^1\alpha_2^1\alpha_2^2\alpha_3^2}{(\alpha_1^2)^2(\alpha_3^1)^2}\right)x_1\nonumber\\
c(1,2)&=&-\frac{(\alpha_1^1)^2\alpha_1^2}{\alpha_2^2\alpha_3^1}+\frac{(\alpha_2^1)^2\alpha_2^2}{\alpha_1^2\alpha_3^1}\nonumber\\
d(1,2)&=&\left(\frac{\alpha_1^1\alpha_3^2}{\alpha_2^1\alpha_2^2}-\frac{(\alpha_1^1)^2(\alpha_3^2)^2}{\alpha_1^2(\alpha_2^2)^2\alpha_3^1}\right)x_2+\left(-\frac{\alpha_2^1\alpha_3^2}{\alpha_1^1\alpha_1^2}+\frac{(\alpha_2^1)^2(\alpha_3^2)^2}{(\alpha_1^2)^2\alpha_2^2\alpha_3^1}\right)x_1
\end{eqnarray}
For the $3/4$ fermion we find:
\begin{equation}
\tilde{\psi}^{(0)}_{\mathcal{L}_1\mathcal{S}_2}=\left(\begin{array}{cc}
\tilde{a}(1,2)&\tilde{b}(1,2)\\
\tilde{c}(1,2)&\tilde{d}(1,2)
\end{array}\right)\qquad
\tilde{\psi}^{(1)}_{\mathcal{L}_1\mathcal{S}_2}=\left(\begin{array}{cc}
\frac{1}{\alpha_3^1\alpha_3^2}\tilde{d}(1,2)&-\frac{\alpha_3^2}{\alpha_3^1}\tilde{c}(1,2)\\
\frac{\alpha_3^1}{\alpha_3^2}\tilde{b}(1,2)&-{\alpha_3^1\alpha_3^2}\tilde{a}(1,2)
\end{array}\right),
\end{equation}
where
\begin{eqnarray}
\tilde{a}(1,2)&=&\left(-\frac{(\alpha_2^1)^2}{\alpha_1^2\alpha_3^1}+\frac{\alpha_2^1\alpha_2^2}{\alpha_1^1\alpha_3^2}\right)x_1\nonumber\\
\tilde{b}(1,2)&=&\left(-\frac{\alpha_1^2(\alpha_2^1)^2}{\alpha_2^2\alpha_3^1}+\frac{(\alpha_2^1)^4\alpha_2^2}{(\alpha_1^1)^2\alpha_1^2\alpha_3^1}\right)x_1x_2+\left(\frac{(\alpha_1^1)^2}{\alpha_3^1}-\frac{(\alpha_2^1)^2(\alpha_2^2)^2}{(\alpha_1^2)^2\alpha_3^1}\right)x_1^2\nonumber\\
\tilde{c}(1,2)&=&\left(-\frac{\alpha_1^1\alpha_1^2\alpha_2^1}{\alpha_3^2}+\frac{(\alpha_2^1)^3(\alpha_2^2)^2}{\alpha_1^1\alpha_1^2\alpha_3^2}\right)x_1\nonumber\\
\tilde{d}(1,2)&=&\left(-\frac{(\alpha_2^1)^2\alpha_2^2\alpha_3^1}{(\alpha_1^1)^2\alpha_1^2}+\frac{(\alpha_2^1)^3\alpha_3^2}{\alpha_1^1(\alpha_1^2)^2}\right)x_1^2+\left(\alpha_3^1-\frac{\alpha_1^1\alpha_2^1\alpha_3^2}{\alpha_1^2\alpha_2^2}\right)x_1x_2
\end{eqnarray}
The charge $1/4$ boson is given by the following expression:
\begin{equation}
\phi^{(0)}_{\mathcal{L}_1\mathcal{S}_2}=\left(\begin{array}{cc}
a(1,2)&b(1,2)\\
c(1,2)&d(1,2)
\end{array}\right)\qquad
\phi^{(1)}_{\mathcal{L}_1\mathcal{S}_2}=\left(\begin{array}{cc}
-\frac{1}{\alpha_3^1\alpha_3^2}d(1,2)&\frac{\alpha_3^2}{\alpha_3^1}c(1,2)\\
-\frac{\alpha_3^1}{\alpha_3^2}b(1,2)&{\alpha_3^1\alpha_3^2}a(1,2)
\end{array}\right),
\end{equation}
where
\begin{eqnarray}
a(1,2)&=&\left(\frac{1}{(\alpha_1^2)^2\alpha_2^2\alpha_3^1}-\frac{1}{\alpha_1^1\alpha_1^2\alpha_2^1\alpha_3^2}\right)x_2+\left(-\frac{(\alpha_1^1)^2}{(\alpha_1^2)^3(\alpha_2^1)^2\alpha_3^1}+\frac{\alpha_1^1\alpha_2^2}{(\alpha_1^2)^2(\alpha_2^1)^3\alpha_3^2}\right)x_1\nonumber\\
b(1,2)&=&-\frac{1}{\alpha_3^1}+\frac{(\alpha_1^1)^2(\alpha_2^2)^2}{(\alpha_1^2)^2(\alpha_2^1)^2\alpha_3^1}\nonumber\\
c(1,2)&=&\left(-\frac{\alpha_1^1}{\alpha_2^1\alpha_2^2\alpha_3^2}+\frac{(\alpha_1^1)^3\alpha_2^2}{(\alpha_1^2)^2(\alpha_2^1)^3\alpha_3^2}\right)x_2+\left(\frac{\alpha_2^1}{\alpha_1^1\alpha_1^2\alpha_3^2}-\frac{\alpha_1^1(\alpha_2^2)^2}{(\alpha_1^2)^3\alpha_2^1\alpha_3^2}\right)x_1\nonumber\\
d(1,2)&=&-\frac{\alpha_2^2\alpha_3^1}{\alpha_1^2(\alpha_2^1)^2}+\frac{\alpha_1^1\alpha_3^2}{(\alpha_1^2)^2\alpha_2^1}
\end{eqnarray}
Finally, the charge $3/4$ boson is:
\begin{equation}
\tilde{\phi}^{(0)}_{\mathcal{L}_1\mathcal{S}_2}=\left(\begin{array}{cc}
\tilde{a}(1,2)&\tilde{b}(1,2)\\
\tilde{c}(1,2)&\tilde{d}(1,2)
\end{array}\right)\qquad
\tilde{\phi}^{(1)}_{\mathcal{L}_1\mathcal{S}_2}=\left(\begin{array}{cc}
-\frac{1}{\alpha_3^1\alpha_3^2}\tilde{d}(1,2)&\frac{\alpha_3^2}{\alpha_3^1}\tilde{c}(1,2)\\
-\frac{\alpha_3^1}{\alpha_3^2}\tilde{b}(1,2)&{\alpha_3^1\alpha_3^2}\tilde{a}(1,2)\end{array}\right),
\end{equation}
where
\begin{eqnarray}
\tilde{a}(1,2)&=&\left(-\frac{(\alpha_1^1)^2\alpha_2^2}{\alpha_1^2(\alpha_2^1)^2\alpha_3^1}+\frac{(\alpha_1^1)^3\alpha_3^2}{(\alpha_1^2)^2\alpha_2^1(\alpha_3^1)^2}\right)x_1^2+\left(\frac{1}{\alpha_3^1}-\frac{\alpha_1^1\alpha_2^1\alpha_3^2}{\alpha_1^2\alpha_2^2(\alpha_3^1)^2}\right)x_1x_2\nonumber\\
\tilde{b}(1,2)&=&\left(\frac{\alpha_1^1\alpha_1^2\alpha_2^1\alpha_3^2}{(\alpha_3^1)^2}-\frac{(\alpha_1^1)^3(\alpha_2^2)^2\alpha_3^2}{\alpha_1^2\alpha_2^1(\alpha_3^1)^2}\right)x_1\nonumber\\
\tilde{c}(1,2)&=&\left(\frac{(\alpha_1^1)^2\alpha_1^2}{\alpha_2^2\alpha_3^1}-\frac{(\alpha_1^1)^4\alpha_2^2}{\alpha_1^2(\alpha_2^1)^2\alpha_3^1}\right)x_1x_2+\left(-\frac{(\alpha_2^1)^2}{\alpha_3^1}+\frac{(\alpha_1^1)^2(\alpha_2^2)^2}{(\alpha_1^2)^2\alpha_3^1}\right)x_1^2\nonumber\\
\tilde{d}(1,2)&=&\left(\frac{\alpha_1^1\alpha_2^2\alpha_3^2}{\alpha_2^1}-\frac{(\alpha_1^1)^2(\alpha_3^2)^2}{\alpha_1^2\alpha_3^1}\right)x_1
\end{eqnarray}
\section{Theta functions}
In this appendix we collect definitions and useful identities for theta functions. Standard references are for instance \cite{Mumford1,FarkasKra}.
The theta functions with characteristics are defined as follows:
 \begin{equation}
\Theta\left[\begin{array}{c}c_1\\c_2\end{array}\right](u,\tau)=\sum_{m\in\mathbb{Z}}q^{(m+c_1)^2/2}e^{2\pi i(u+c_2)(m+c_1)},
\end{equation}
where $q=e^{2\pi i \tau}$. \\
For our purpose we need the Jacobi theta functions:
\begin{equation}
\Theta_1(u,\tau)\equiv\Theta\left[\begin{array}{c}\frac{1}{2}\\\frac{1}{2}\end{array}\right](u,\tau)\qquad \Theta_2(u,\tau)\equiv\Theta\left[\begin{array}{c}\frac{1}{2}\\0\end{array}\right](u,\tau)
\end{equation}
\begin{equation}
\Theta_3(u,\tau)\equiv\Theta\left[\begin{array}{c}0\\0\end{array}\right](u,\tau)\qquad\Theta_4(u,\tau)\equiv\Theta\left[\begin{array}{c}0\\\frac{1}{2}\end{array}\right](u,\tau)
\end{equation}
We write $\Theta_i(0,\tau)\equiv \Theta_i(\tau)$.\\
These theta functions are symmetric in the $u$--argument:
\begin{equation}
\Theta_1(-u,\tau)=-\Theta_1(u,\tau)\quad\Theta_2(-u,\tau)=\Theta_2(u,\tau)\quad\Theta_3(-u,\tau)=\Theta_3(u,\tau)\quad\Theta_4(-u,\tau)=\Theta_4(u,\tau)\nonumber
\end{equation}
In particular, one sees that $\Theta_1(0,\tau)=0$.
To uniformize the $\alpha_i$ (\ref{uniformization}), we used the identities
\begin{eqnarray}
\Theta_3^2(u,\tau)\Theta_4^2(\tau)&=&\Theta_4^2(u,\tau)\Theta_3^2(\tau)-\Theta_1^2(u,\tau)\Theta_2^2(\tau)\nonumber\\
\Theta_2^2(u,\tau)\Theta_4^2(\tau)&=&\Theta_4^2(u,\tau)\Theta_2^2(\tau)-\Theta_1^2(u,\tau)\Theta_3^2(\tau).
\end{eqnarray}
In order to simplify the cohomology elements we applied the following addition rules:
\begin{align}
\label{simp1}
\Theta_4(u_1+u_2,\tau)\Theta_4(u_1-u_2,\tau)\Theta_4(0,\tau)^2&=\Theta_4(u_1,\tau)^2\Theta_4(u_2,\tau)^2-\Theta_1(u_1,\tau)^2\Theta_1(u_2,\tau)^2\nonumber\\
\Theta_1(u_1+u_2,\tau)\Theta_1(u_1-u_2,\tau)\Theta_3(0,\tau)^2&=\Theta_1(u_1,\tau)^2\Theta_3(u_2,\tau)^2-\Theta_3(u_1,\tau)^2\Theta_1(u_2,\tau)^2\nonumber\\
&=\Theta_4(u_1,\tau)^2\Theta_2(u_2,\tau)^2-\Theta_2(u_1,\tau)^2\Theta_4(u_2,\tau)^2\nonumber\\
\Theta_1(u_1+u_2,\tau)\Theta_2(u_1-u_2,\tau)\Theta_4(0,\tau)^2&=\Theta_1(u_1,\tau)^2\Theta_4(u_2,\tau)^2-\Theta_4(u_1,\tau)^2\Theta_1(u_2,\tau)^2
\end{align}
\begin{align}
\label{simp2}
\Theta_1(u_1+u_2,\tau)\Theta_4(u_1-u_2,\tau)\Theta_2(0,\tau)\Theta_3(0,\tau)&=\Theta_1(u_1,\tau)\Theta_2(u_2,\tau)\Theta_3(u_2,\tau)\Theta_4(u_1,\tau)+(u_1\leftrightarrow u_2)\nonumber\\
\Theta_4(u_1+u_2,\tau)\Theta_1(u_1-u_2,\tau)\Theta_2(0,\tau)\Theta_3(0,\tau)&=\Theta_1(u_1,\tau)\Theta_2(u_2,\tau)\Theta_3(u_2,\tau)\Theta_4(u_1,\tau)-(u_1\leftrightarrow u_2)\
\end{align}
These identities are actually just special cases of a more general identities. In order to determine the correlators we need the most general addition theorems. 
For this, we introduce some more notation \cite{Mumford1}:
\begin{equation}
x_1=\frac{1}{2}(x+y+u+v)\quad y_1=\frac{1}{2}(x+y-u-v)\quad u_1=\frac{1}{2}(x-y+u-v)\quad v_1=\frac{1}{2}(x-y-u+v)
\end{equation} 
Furthermore we define $\Theta_i^u\equiv\Theta_i(u,\tau)$. For our calculations we can make use of the following formulas:
\begin{align}
\label{bigtheta1}
-\Theta_1^x\Theta_1^y\Theta_1^u\Theta_1^v-\Theta_2^x\Theta_2^y\Theta_2^u\Theta_2^v+\Theta_3^x\Theta_3^y\Theta_3^u\Theta_3^v+\Theta_4^x\Theta_4^y\Theta_4^u\Theta_4^v&=2\Theta_1^{x_1}\Theta_1^{y_1}\Theta_1^{u_1}\Theta_1^{v_1}\nonumber \\
\Theta_1^x\Theta_1^y\Theta_1^u\Theta_1^v-\Theta_2^x\Theta_2^y\Theta_2^u\Theta_2^v+\Theta_3^x\Theta_3^y\Theta_3^u\Theta_3^v-\Theta_4^x\Theta_4^y\Theta_4^u\Theta_4^v&=2\Theta_4^{x_1}\Theta_4^{y_1}\Theta_4^{u_1}\Theta_4^{v_1}
\end{align}
\begin{align}
\label{bigtheta2}
\Theta_3^x\Theta_3^y\Theta_2^u\Theta_2^v+\Theta_4^x\Theta_4^y\Theta_1^u\Theta_1^v-\Theta_2^x\Theta_2^y\Theta_3^u\Theta_3^v-\Theta_1^x\Theta_1^y\Theta_4^u\Theta_4^v&=2\Theta_4^{x_1}\Theta_4^{y_1}\Theta_1^{u_1}\Theta_1^{v_1}\nonumber \\
\Theta_3^x\Theta_3^y\Theta_2^u\Theta_2^v-\Theta_4^x\Theta_4^y\Theta_1^u\Theta_1^v-\Theta_2^x\Theta_2^y\Theta_3^u\Theta_3^v+\Theta_1^x\Theta_1^y\Theta_4^u\Theta_4^v&=2\Theta_1^{x_1}\Theta_1^{y_1}\Theta_4^{u_1}\Theta_4^{v_1}
\end{align}
What we need for our calculations are the differences of the two relations (\ref{bigtheta1}) and (\ref{bigtheta2}):
\begin{align}
\label{mytheta}
\Theta_1^x\Theta_1^y\Theta_1^u\Theta_1^v-\Theta_4^x\Theta_4^y\Theta_4^u\Theta_4^v&=\Theta_4^{x_1}\Theta_4^{y_1}\Theta_4^{u_1}\Theta_4^{v_1}-\Theta_1^{x_1}\Theta_1^{y_1}\Theta_1^{u_1}\Theta_1^{v_1}\nonumber\\
\Theta_1^x\Theta_1^y\Theta_4^u\Theta_4^v-\Theta_4^x\Theta_4^y\Theta_1^u\Theta_1^v&=\Theta_1^{x_1}\Theta_1^{y_1}\Theta_4^{u_1}\Theta_4^{v_1}-\Theta_4^{x_1}\Theta_4^{y_1}\Theta_1^{u_1}\Theta_1^{v_1}
\end{align}

\providecommand{\href}[2]{#2}\begingroup\raggedright\endgroup

\end{document}